\documentclass[prd,showpacs,preprintnumbers]{revtex4}

\usepackage{graphicx}
\usepackage{dcolumn}
\usepackage{bm}
\usepackage{latexsym}
\usepackage{amsfonts}
\usepackage{amssymb}
\newcommand{\R }{ \mathop{R^{\mu}_{\ \nu}}^{(4)}  }
\newcommand{\Rs }{ \mathop{R}^{(4)}  }

\begin{document}

\preprint{KUCP0213 }

\title{ Radion and Holographic Brane Gravity }

\author{Sugumi Kanno}
\email{kanno@phys.h.kyoto-u.ac.jp}
 \affiliation{%
 Graduate School of Human and Environment Studies, Kyoto University, Kyoto
 606-8501, Japan 
}%

\author{Jiro Soda}
\email{jiro@phys.h.kyoto-u.ac.jp}
\affiliation{
 Department of Fundamental Sciences, FIHS, 
Kyoto University, Kyoto 606-8501, Japan
}%

\date{\today}

\begin{abstract}
The low energy effective theory for the Randall-Sundrum two brane system 
is investigated with an emphasis on the role of the non-linear radion 
in the brane world.  The equations of motion in the bulk is solved 
using a low energy expansion method. 
This allows us, through the junction conditions, 
to deduce the effective equations of motion for the gravity on the brane. 
It is  shown that the gravity on the brane world is described by 
a quasi-scalar-tensor theory with a specific coupling function 
$\omega(\Psi) = 3\Psi /2(1-\Psi ) $ on the positive tension brane 
and $\omega(\Phi) = -3\Phi /2(1+\Phi ) $ on the negative tension brane,
where $\Psi$ and $\Phi$ are non-linear realizations of the radion
on the positive and negative tension branes, respectively. 
In contrast to the usual scalar-tensor gravity, the quasi-scalar-tensor 
gravity couples with two kinds of matter, namely, the matters
on both positive and negative tension branes, with different 
effective gravitational coupling constants.
  In particular, 
the radion disguised as the scalar fields $\Psi$ and $\Phi$ couples with 
the sum of the traces of the energy momentum tensor on both branes. 
 In the course of the derivation, it has been revealed 
that the radion plays an essential role to convert
 the non-local Einstein gravity with the 
generalized dark radiation to the local quasi-scalar-tensor gravity. 
For completeness, we also derive the effective action for our theory
by substituting the bulk solution into the original action.  
It is also shown that the quasi-scalar-tensor gravity works as  holograms 
at the low energy in the sense that the bulk geometry can be reconstructed 
from the solution of the quasi-scalar-tensor gravity. 
\end{abstract}

\maketitle

\section{Introduction}

Motivated by the recent development of the superstring theory, 
the brane world scenario has been studied intensively. 
In particular, the warped compactification mechanism proposed by 
Randall and Sundrum has given birth to a new picture of the 
universe~\cite{RS}. The single brane model (RS2) is well studied so far 
because of its simplicity and the absence of the stability problem of 
the radion mode~\cite{BWC1,BWC2,KJ,GKR}. 
As for the two brane model (RS1),  
Garriga and Tanaka have shown that the gravity  on the brane 
behaves as the Brans-Dicke theory at the  linearized level~\cite{GT}.
 Thus, the conventional linearized Einstein equations 
do not hold even on scales large compared with the curvature scale $l$ 
in the bulk. Charmousis et al. have clearly identified the Brans-Dicke field 
as the radion mode~\cite{CR}. The subsequent research  has been 
focussed  on the role of the radion in the brane world 
scenario~\cite{gen,chiba,low} . 
 
However, the above-mentioned works are restricted to the linear theory 
or to homogeneous cosmological models.  
To study the non-linear gravity is important for  applications 
to astrophysical and cosmological problems. 
Recently, Wiseman have analyzed a special two brane system  
with the  negative tension brane taken to be in vacuum 
and shown that the low energy effective theory becomes the 
scalar-tensor theory with a specific coupling function~\cite{wiseman}. 
Here, we consider the general case including the matter on 
the negative tension brane and derive the effective equations of motion 
for this system using a low energy expansion method 
developed by us~\cite{kanno}.      

To further illuminate the role of the radion in the brane world, 
let us pose  the issue in the following way.   In our previous 
paper, we have derived the low energy effective  equation on the brane 
as~\cite{kanno} (see also~\cite{SMS})
\begin{equation}
     	G^\mu_{\ \nu}  = {\kappa^2 \over l} T^{\mu}_{\ \nu} 
        	-{2\over l} \chi^\mu_{\ \nu} (x^\mu ) \ ,
\label{loweq}
\end{equation}
where $G^\mu_{\ \nu}$, $\kappa$ and $T^{\mu}_{\ \nu} $ 
denote the 4-dimensional Eisnstein tensor, the 5-dimensional gravitational 
constant, and the energy momentum tensor on the brane, respectively. 
Here,  the `` constant of integration " 
$\chi_{\mu\nu} (x) $   is transverse and traceless.  
When we impose the maximal symmetry on the spatial part of the brane world, 
Eq.~({\ref{loweq}) reduces to the  
Friedmann equation with the dark radiation 
\begin{equation}
   	H^2 = {8\pi G\rho \over 3}  + {{\cal C} \over a_0^4}
\label{cosmoeq}
\end{equation}
where $H$, $a_0$ and $\rho$ are, respectively, the Hubble parameter, 
the scale factor and the total energy density of each brane, while 
${\cal C}$ is a constant of integration associated with the mass of
a black hole in the bulk. 
Hence,   $\chi_{\mu\nu} (x) $  can be regarded as a generalization 
of the  dark  radiation appeared in Eq.~(\ref{cosmoeq}).
The point is that 
Eq.~(\ref{loweq})  holds irrespective of the existence of other branes.
The effect of the bulk geometry comes in to
the brane world  only through $\chi_{\mu\nu}$.

On the other hand, as we have noted, the scalar-tensor theory emerges 
in the two brane system. 
How can we reconcile these seemingly incompatible pictures? 
In this paper, we reveal a  mechanism to convert the 
Einstein equations with the generalized dark radiation 
to the quasi-scalar-tensor gravity.  After all, it turns out that the radion 
disentangles the non-locality in the non-conventional Einstein 
equations and leads to the local quasi-scalar-tensor gravity. 

This paper is organized as follows. In section 2, our iteration 
scheme to solve the Einstein equations at the low energy is explained. 
In section 3, the background solution is presented. 
In section 4, we derive the brane effective action from the junction 
conditions at the leading order. We see the effective theory is described by 
the quasi-scalar-tensor gravity with a specific coupling function. 
The relation to the holography is also discussed. 
In section 5, a systematic method to compute 
the higher order corrections is discussed. 
Section 6 is devoted to discussions and conclusion. 
In Appendix A, we explain the physical meaning of our method,
especially the relation to the zero mode and Kaluza-Klein modes
in the linear theory, by using a simple scalar field model. 
In Appendix B, the linearized gravity is analyzed using our method in 
detail. 

\section{Low  Energy  Approximation}
 
\subsection{RS1 Model and Basic Equations}  
   
The model is described by the action 
\begin{equation}
 S = {1\over 2\kappa^2}\int d^5 x \sqrt{-g} \left( 
 {\cal R} + {12\over l^2} \right)  
 -\sum_{i=A,B} \sigma_i \int d^4 x \sqrt{-g^{i{\rm brane}}}
 + \sum_{i=A,B} \int d^4 x \sqrt{-g^{i{\rm brane}}} 
 {\cal L}_{\rm matter}^i \ ,
 \label{eq:action}
\end{equation}
where ${\cal R}$, $g^{i{\rm brane}}_{\mu\nu}$ and $\kappa^2  $  are 
the scalar curvature, the induced metric on  branes and the
gravitational constant in 5-dimensions, respectively. We  consider 
an $S_1/Z_2$ orbifold spacetime with the two branes as the fixed points.
In the RS1 model, the two flat 3-branes 
are embedded in the 5-dimensional asymptotically anti-deSitter (AdS) bulk 
with the curvature radius $l$ with brane tensions given by
$\sigma_A=6/(\kappa^2l)$ and $\sigma_B=-6/(\kappa^2l)$.

For general non-flat branes, we can not keep both of the two branes
straight in the Gaussian normal coordinate system.
Hence, we use the following coordinate system to describe 
the geometry of the brane model;
\begin{equation}
ds^2 = e^{2\phi (y, x^{\mu})} dy^2 
+ g_{\mu\nu} (y,x^{\mu} ) dx^{\mu} dx^{\nu}  \ .
\label{eq:5-dim-metric}
\end{equation}
\begin{figure}[h]
\centerline{\includegraphics[width = 6cm, height = 7cm]{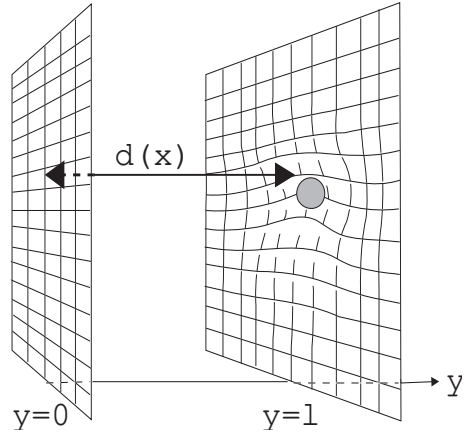}}
\caption{Radion as a distance between two branes.}
\end{figure}
We place the branes  at $y=0$ ($A$-brane) and $y=l$ ($B$-brane) 
in this coordinate system. The proper distance between
the two branes
 with fixed $x$ coordinates can be written as
\begin{equation}
   	d(x) = \int_0^l e^{\phi (y,x)} dy  \ .
   	\label{eq:distande}
\end{equation}
Hence, we call $\phi$ the radion. 
In this coordinate system, the 5-dimensional Einstein equations 
become
\begin{eqnarray}
 	& & e^{-\phi} (e^{-\phi} K^\mu_{\ \nu })_{,y} 
        	- (e^{-\phi} K)(e^{-\phi} K^\mu_{\ \nu}) 
        	+ {\R} -\nabla^\mu \nabla_\nu \phi 
        	-\nabla^\mu \phi  \nabla_\nu \phi \nonumber \\
        	&& \qquad = - {4\over l^2} \delta^\mu_\nu  
        	+ \kappa^2 \left({1\over 3} \delta^\mu_\nu \sigma_A
        	+T^{A\mu}_{\quad \nu} -{1\over 3} \delta^\mu_\nu T^A \right)
        	e^{-\phi} \delta (y) 
        	+ \kappa^2 \left({1\over 3} \delta^\mu_\nu \sigma_B
        	+\tilde{T}^{B\mu}_{\quad \nu} -{1\over 3} \delta^\mu_\nu 
        	\tilde{T}^B 
        	\right)	e^{-\phi} \delta (y-l)   \ ,
        	\label{eq:k-munu} \\	
 	& & e^{-\phi} (e^{-\phi} K )_{, y} 
       		- (e^{-\phi} K^{\alpha  \beta})(e^{-\phi} K_{\alpha  \beta} )
       		- \nabla^\alpha \nabla_\alpha \phi 
       		-\nabla^\alpha \phi \nabla_\alpha \phi \nonumber \\
        	&&\qquad     = - {4\over l^2}  
     		-{\kappa^2 \over 3} \left(-4\sigma_A +T^A \right) 
        	e^{-\phi}\delta (y)  
     		-{\kappa^2 \over 3} \left(-4\sigma_B +\tilde{T}^B \right) 
        	e^{-\phi}\delta (y-l)  \ , 
        	\label{eq:k-yy}	\\
 	& & \nabla_\nu (e^{-\phi} K_{\mu }^{\  \nu})  
        	- \nabla_\mu (e^{-\phi} K )   =  0  \ ,
        	\label{eq:k-ymu}
\end{eqnarray}
where $\R$ is the curvature on the brane and $\nabla_\mu $ denotes the
covariant derivative with respect to the metric $g_{\mu\nu}$.    
And we introduced the tensor 
$K_{\mu\nu} = -g_{\mu\nu,y}/2$ for convenience. 
One can read off the junction condition from the above equations as
\begin{eqnarray}
 	e^{-\phi} \left[ K^\mu_\nu - \delta^\mu_\nu K \right] |_{y=0} 
    		&=& {\kappa^2 \over 2}  
       		\left( -\sigma_A \delta^\mu_\nu 
       		+ T^{A\mu}_{\quad\nu} \right) \ ,
       		\label{JC-k:A}  \\	
 	e^{-\phi} \left[ K^\mu_\nu - \delta^\mu_\nu K \right] |_{y=l} 
    		&=& -{\kappa^2 \over 2}  
        	\left( -\sigma_B \delta^\mu_\nu 
        	+ \tilde{T}^{B\mu}_{\quad\nu} \right) \ ,
        	\label{JC-k:B}	
\end{eqnarray}
where $K^\mu_\nu=g^{\mu\alpha}K_{\alpha\nu}$
and the fact that we are considering a $Z_2$ symmetric spacetime
is used. 
Decompose the extrinsic curvature into the traceless part and the trace part
\begin{equation}
  	e^{-\phi}K_{\mu\nu} 
    		= \Sigma_{\mu\nu} + {1\over 4} g_{\mu\nu} Q  \ , \quad
    	Q = - e^{-\phi}{\partial \over \partial y}\log \sqrt{-g}    \  .
    	\label{eq:decompose} 
\end{equation}
Then, off the brane, we obtain the basic equations;
\begin{eqnarray}
 	& & e^{-\phi} \Sigma^\mu_{\ \nu , y} - Q \Sigma^\mu_{\ \nu} 
    		= -\left[ \R - {1\over 4} \delta^\mu_\nu \Rs 
            	-\nabla^\mu \nabla_\nu \phi 
               	-\nabla^\mu \phi  \nabla_\nu \phi
               	+{1\over 4} \delta^\mu_\nu 
               	\left( \nabla^\alpha \nabla_\alpha \phi 
       		+\nabla^\alpha \phi \nabla_\alpha \phi \right)
       		\right]      \ ,     
       		\label{eq:munu-traceless} \\
	& & {3\over 4} Q^2 - \Sigma^\alpha_{\ \beta} \Sigma^\beta_{\ \alpha} 
    		= \left[ \Rs \right] + {12\over l^2}   \ ,  
    		\label{eq:munu-trace} \\
 	& & e^{-\phi}  Q_{, y} -{1\over 4}Q^2 
       		- \Sigma^{\alpha \beta} \Sigma_{\alpha \beta} 
    		= \nabla^\alpha \nabla_\alpha \phi 
       		+ \nabla^\alpha \phi \nabla_\alpha \phi - {4\over l^2}  \ , 
       		\label{eq:yy}  \\
 	& & \nabla_\lambda \Sigma_{\mu }^{\ \lambda}  
        	- {3\over 4} \nabla_\mu Q = 0   \ .
        	\label{eq:ymu}
\end{eqnarray}
And the junction conditions are
\begin{eqnarray}
   	\left[ \Sigma^{\mu}_{\nu} 
   		- {3\over 4} \delta^\mu_\nu Q \right] \Bigg|_{y=0}
    		&=& {\kappa^2 \over 2} (-\sigma_A \delta^\mu_\nu 
    		+ T^{A\mu}_{\quad\nu})  \ , 
    		\label{JC:A} \\
    	\left[ \Sigma^{\mu}_{\nu} 
   		- {3\over 4} \delta^\mu_\nu Q \right] \Bigg|_{y=l}
    		&=& -{\kappa^2 \over 2} (-\sigma_B \delta^\mu_\nu + 
    		\tilde{T}^{B\mu}_{\quad\nu})    \ .
    		\label{JC:B}
\end{eqnarray}
The problem now is separated into two parts. First, we must solve 
the bulk equations of motion with the Dirichlet boundary condition 
at the A-brane, $g_{\mu\nu} (y=0 ,x^\mu ) = h_{\mu\nu} (x^\mu ) $. 
Then, the junction condition is imposed at each brane. 
As the junction conditions constrain the induced metrics on both
branes, they naturally give rise to the effective equations of motion 
for the gravity on the branes.

\subsection{Low Energy Expansion Scheme}

Unfortunately,  it is a formidable task to solve the 5-dimensional 
Einstein equations exactly. However, notice that typically the length 
scale of the internal space is $l\ll 0.1 $ mm. On the other hand, 
usual astrophysical and cosmological phenomena take place at scales 
much larger than this scale. 
Then we need only the low energy effective theory to analyze the variety 
of problems, for example,  the formation of a black hole, the propagation 
of gravitational waves, the evolution of cosmological perturbations, 
and so on. It should be stressed that the low energy does not necessarily 
implies weak gravity on the branes. 

Along the normal coordinate $y$, the metric varies with the characteristic 
length scale $l$; $ g_{\mu\nu ,y} \sim g_{\mu\nu} /l$.  
Denote the characteristic 
length scale of the curvature  on the brane as $L$.
Then we have
$ R \sim g_{\mu\nu} / L^2 $. For reader's reference, let us take 
$l=1$ mm, for example. Then, the relations in the RS1 model 
\begin{equation}
 	\kappa^2 \sigma_A = {6\over l} \ , \qquad 
 	{\kappa^2 \over l} = 8\pi G_N  
 	\label{rel:RS1}
\end{equation}
give $\kappa^2 \sim (10^8 {\rm GeV})^{-3} $ 
and $|\sigma_i|\sim 1 {\rm TeV}^4 $.  

In this paper, we will consider the low energy 
regime in the sense that the energy density of the matter, $\rho_i$, 
on a brane is smaller than the brane tension, i.e.,
$\rho_i /|\sigma_i| \ll 1$. 
In this regime, a simple dimensional analysis,
\begin{equation}
{\rho_i \over |\sigma_i|} 
	\sim {l \over \kappa^2 |\sigma_i|}\,{\kappa^2\rho_i\over l}
	\sim ({l\over L})^2 \ll 1 
	\label{rel:low-energy}
\end{equation}
implies that the curvature on the brane can be neglected compared with the 
extrinsic curvature at low energies. 
Thus, the Anti-Newtonian or gradient expansion method used in the cosmological 
context~\cite{tomita} is applicable to our problem. 

Our iteration scheme is to write the metric $g_{\mu\nu}$ 
as a sum of local tensors built out of the induced metric on the 
brane, with the number of derivatives increasing with the order
of iteration, that is, 
$ O((l/L)^{2n})$, $n=0,1,2,\cdots$. 
Hence, we seek the metric as a perturbative series 
\begin{eqnarray}
    	&&  g_{\mu\nu} (y,x^\mu ) =
  		a^2 (y,x) \left[ h_{\mu\nu} (x^\mu) 
  		+ g^{(1)}_{\mu\nu} (y,x^\mu)
      		+ g^{(2)}_{\mu\nu} (y, x^\mu ) + \cdots  \right]  \ , 
      		\label{expansion:metric} \\
   	&&  g^{(n)}_{\mu\nu} (y=0 ,x^\mu ) =  0    \ , \quad n=1,2,3,...
   		\label{BC}
\end{eqnarray}
where the factor $a^2 (y,x) $ is extracted because of the reason 
explained later and we put the Dirichlet boundary condition 
$g_{\mu\nu} (y=0, x) =  h_{\mu\nu} (x)$ at the $A$-brane. 
We do not need to know the geometry of the $B$-brane 
when we focus on the effective 
equations on the $A$-brane. In other words, from the viewpoint on the
$A$-brane, the junction condition at the $B$-brane simply gives the 
boundary condition for the bulk geometry. 
Other quantities are also expanded as 
\begin{equation}
   	\Sigma^\mu_{\ \nu} = \Sigma^{(0)\mu}_{\quad\ \nu}
            	+\Sigma^{(1)\mu}_{\quad\ \nu}+\Sigma^{(2)\mu}_{\quad\ \nu}
                +\cdots  \ .
                \label{expansion:sigma}
\end{equation}
In Appendix A, we illustrate our method using a simple scalar field
example to clarify the relation of the low energy expansion with
the zero mode and Kaluza-Klein modes in the linearized theory. 

\section{Background Geometry}

As we can ignore the matter at the lowest order, we obtain the vacuum 
brane. Namely, we have an almost flat brane compared with the curvature scale 
of the bulk space-time. 
At the 0-th order,  we can neglect the curvature term. Then, we have 
\begin{eqnarray}
 	& & e^{-\phi}\Sigma^{(0)\mu}_{\quad \  \nu , y} 
      		- Q^{(0)} \Sigma^{(0)\mu}_{\quad \  \nu} 
    		= 0 \ ,     
    		\label{0:munu-traceless}  \\
 	& &   {3\over 4} Q^{(0)2} - \Sigma^{(0)\alpha}_{\quad \  \beta} 
    		\Sigma^{(0)\beta}_{\quad \  \alpha} 
    		=  {12\over l^2} \ ,  
    		\label{0:munu-trace}   \\
 	& &  e^{-\phi} Q^{(0)}_{, y} -{1\over 4}Q^{(0)2} 
    		- \Sigma^{(0)\alpha \beta} \Sigma^{(0)}_{\alpha \beta} 
    		= - {4\over l^2}  \ ,    
    		\label{0:yy}   \\
 	& &   \nabla_\lambda \Sigma^{(0)\ \lambda}_{\quad \ \mu }   
    		-  {3\over 4} \nabla_\mu Q^{(0)} = 0  \ .
    		\label{0:ymu} 
\end{eqnarray}
The junction condition is
\begin{eqnarray}
   	\left[ \Sigma^{(0)\mu}_{\quad\ \nu} 
   		- {3\over 4} \delta^\mu_\nu Q^{(0)} \right] \Bigg|_{y=0}
    		&=& - {\kappa^2 \over 2} \sigma_A \delta^\mu_\nu  \ ,
    		\label{0:JC-A}  \\
    	\left[ \Sigma^{(0)\mu}_{\quad\ \nu} 
   		- {3\over 4} \delta^\mu_\nu Q^{(0)} \right] \Bigg|_{y=l}
    		&=& {\kappa^2 \over 2} \sigma_B \delta^\mu_\nu  \ .
    		\label{0:JC-B}
\end{eqnarray}
Using Eq.~(\ref{eq:decompose}), Eq.~(\ref{0:munu-traceless}) can be readily 
integrated,
\begin{equation}
	\Sigma^{(0)\mu}_{\quad \  \nu} 
    		= {C^\mu_{\ \nu} (x^\mu) \over \sqrt{-g} } 
        	\ , \qquad C^\mu_{\ \mu} =0  \ ,
        	\label{0:sigma}
\end{equation}
where $C^\mu_{\ \nu}$ is a ``constant" of integration. 
This term is not allowed to exist because of the junction conditions
(\ref{0:JC-A}) and (\ref{0:JC-B}). 
Then, it is easy to solve the remaining equations. The result is
\begin{equation}
    	\Sigma^{(0)\mu}_{\quad \  \nu} =0 \ , \qquad  Q^{(0)} = {4\over l} \ .
    	\label{0:Q}
\end{equation}
Using the definition 
\begin{equation}
     	K^{(0)}_{\mu\nu}  
     		= - {1\over 2} {\partial \over \partial y} 
        	g^{(0)}_{\mu\nu}  = \frac{1}{l}e^{\phi}g^{(0)}_{\mu\nu}   \   ,
        	\label{0:k}
\end{equation}
we get the 0-th order metric  as
\begin{equation}
 	ds^2 = e^{2\phi (y,x) } dy^2 + a^2 (y,x)
        	h_{\mu\nu}(x^\mu ) dx^\mu dx^\nu  \ , \qquad
      	a(y,x) = \exp\left[-{1\over l} \int^y_0 dy e^{\phi (y,x)}\right] \ , 
      	\label{0:metric}
\end{equation}
where  the tensor $h_{\mu\nu}$ is  the induced metric on the positive tension 
brane.  Note that the metric  derived by Charmousis et al.,
$e^\phi = 1+ 2 f(x) e^{2y/l } /l $,
 is consistent with this solution~\cite{CR}. To proceed further, we take 
 the coordinate system to be $\phi (y,x ) = \phi (x)$. 
Then, we have $a(y,x) = \exp [-y e^{\phi} /l ]$. 
 Though this choice of the coordinate system is generally possible at least 
 locally, there may be a global obstruction.  
 However, as we show below, we can consistently get non-trivial solutions. 
 Moreover, we explicitly demonstrate the validity of our choice   
 at the level of linear theory in Appendix B. 

 Given the 0-th order solution, junction conditions (\ref{0:JC-A}) and 
 (\ref{0:JC-B})  lead to the well known relations 
\begin{equation}
 	\kappa^2 \sigma_A = \frac{6}{l} \ , \quad \kappa^2 \sigma_B 
 		= -\frac{6}{l} \ .
 		\label{0:JC}
\end{equation}
Note that $\phi(x)$ and $h_{\mu\nu}(x)$ are arbitrary functions of $x$ at 
the 0-th order.

\section{ Holographic Quasi-Scalar-Tensor Gravity }

\subsection{Bulk Geometry }

The next order solution is obtained by taking into account the 
terms neglected at the 0-th order. 
It is at this order that the effect of matter comes in.  
At the 1-st order, Eqs.~(\ref{eq:munu-traceless}-\ref{eq:ymu}) become
\begin{eqnarray}
 	& & e^{-\phi}\Sigma^{(1) \mu}_{\quad \  \nu , y} 
      		- {4\over l} \Sigma^{(1) \mu}_{\quad \  \nu} 
    		= -\left[ \R - \nabla^\mu \nabla_\nu \phi 
       		- \nabla^\mu \phi \nabla_\nu \phi 
       		\right]^{(1)}_{\rm trace less}  \ , 
       		\label{1:munu-traceless} \\
	& & {6 \over l} Q^{(1)}  = \left[\Rs \right]^{(1)}   \  ,
		\label{1:munu-trace} \\
 	& & e^{-\phi} Q^{(1)}_{, y} -{2\over l}Q^{(1)} 
 		= \left[ \nabla^\alpha \nabla_\alpha \phi 
       		+ \nabla^\alpha \phi \nabla_\alpha \phi 
       		\right]^{(1)} \ ,
       		\label{1:yy} \\
 	& & \nabla_\lambda \Sigma_{\quad \mu }^{(1)\ \lambda}   
              	-  {3\over 4} \nabla_\mu Q^{(1)} = 0 \ ,
              	\label{1:ymu}
\end{eqnarray}
where the subscript ``traceless" represents the traceless part of the quantity
in the square brackets.
The junction conditions are given by
\begin{eqnarray}
   	\left[ \Sigma^{(1)\mu}_{\quad\ \nu} 
   		- {3\over 4} \delta^\mu_\nu Q^{(1)} \right] \Bigg|_{y=0}
    		&=& {\kappa^2 \over 2} T^{A\mu}_{\quad\nu}  \ , 
    		\label{1:JC-A} \\
    	\left[ \Sigma^{(1)\mu}_{\quad\ \nu} 
   		- {3\over 4} \delta^\mu_\nu Q^{(1)} \right] \Bigg|_{y=l}
    		&=& -{\kappa^2 \over 2} \tilde{T}^{B\mu}_{\quad\nu}    \ ,
    		\label{1:JC-B}
\end{eqnarray}
where the superscript $(1)$ represents the order of the gradient expansion. 
Here, $[R^{(4)\mu}_{\quad \ \nu} ]^{(1)} $ means the Ricci tensor of 
$a^2 h_{\mu\nu} $.  
Note that now $a=\exp[- y e^{\phi}/l]$. 
It is convenient to introduce the Ricci
tensor of $h_{\mu\nu}$, denoted by $R^\mu_\nu (h)$, and
express $[R^{(4)\mu}_{\quad \ \nu} ]^{(1)} $ in terms of
$R^\mu_\nu$ and $\phi$;  
\begin{equation}
 	\left[ \R (g) \right]^{(1)}
   		= {1\over a^2}\left[  R^\mu_{\ \nu} (h)
  		+ 2{y\over l} e^{\phi} \left(
  		\phi^{|\mu}_{\ |\nu}+\phi^{|\mu} \phi_{|\nu} \right) 
  		+\delta^\mu_\nu {y\over l} e^{\phi} \left(
  		\phi^{|\alpha}_{\ |\alpha}+\phi^{|\alpha} \phi_{|\alpha} 
  		\right) 
  		+ 2{y^2 \over l^2} e^{2\phi} \phi^{|\mu} \phi_{|\nu}
   		- 2\delta^\mu_\nu {y^2 \over l^2} e^{2\phi}
   		\phi^{|\alpha} \phi_{|\alpha} \right]     \ ,
   		\label{1:R}
\end{equation}
where $|$ denotes the covariant derivative with respect to $h_{\mu\nu}$. 
Similarly it is convenient to express the second derivatives of $\phi$
as
\begin{equation}
 	\left[ \nabla^\mu \nabla_\nu \phi \right]^{(1)}= {1\over a^2} \left[
  		\phi^{|\mu}_{\ |\nu} + 2{y\over l}e^{\phi} \phi^{|\mu} 
  		\phi_{|\nu} - {y\over l}e^{\phi} \delta^\mu_\nu \phi^{|\alpha} 
  		\phi_{|\alpha}  \right].  
  		\label{1:phi}
\end{equation}

Substituting the trace of Eq.~(\ref{1:R}) into the right-hand side of 
Eq.~(\ref{1:munu-trace}), 
we obtain
\begin{equation}
   	Q^{(1)} = {l \over a^2} \left[ {1\over 6} R(h) 
      		+ {y e^{\phi} \over l}\left(  \phi^{|\alpha}_{\ |\alpha} 
       		+  \phi^{|\alpha}  \phi_{|\alpha} \right) 
       		- {y^2 e^{2\phi} \over l^2}  \phi^{|\alpha} \phi_{|\alpha} 
       		\right]    \ .
       		\label{1:Q}
\end{equation}
Note that Eq.~(\ref{1:yy}) is trivially satisfied now. 
Hereafter, we omit the argument of the curvature for simplicity. 
Substituting  Eqs.~(\ref{1:R}) and (\ref{1:phi}) into 
Eq.~(\ref{1:munu-traceless}) and integrating it, 
we obtain the traceless part of the extrinsic curvature as 
\begin{equation}
  	\Sigma^{(1)\mu}_{\quad \nu} =  {l \over  a^2 } \left[ 
      		{1\over 2}\left( R^\mu_{\ \nu}  
                - {1\over 4} \delta^\mu_\nu R \right)
                + {y e^{\phi} \over l}\left( \phi^{|\mu}_{\ |\nu}  
       		-{1\over 4}\delta^\mu_\nu  \phi^{|\alpha}_{\ |\alpha} 
        	\right) 
       		+( {y^2 e^{2\phi} \over l^2} + {y e^{\phi} \over l})
       		\left(  \phi^{|\mu} \phi_{|\nu} -{1\over 4}\delta^\mu_\nu
       		\phi^{|\alpha}  \phi_{|\alpha} \right) 
       		\right]   
        	+ {\chi^{\mu}_{\ \nu} (x) \over a^4}  \ ,
        	\label{1:sigma}
\end{equation}
where $\chi^{\mu}_{\nu}$ is an integration constant
with the property $\chi^{\mu}_{\ \mu} =0 $.  And
$\chi^{\mu}_{\nu}$ 
must be transverse  $\chi^{\mu}_{\ \nu|\mu}=0 $ in order to satisfy 
Eq.~(\ref{1:ymu}).  The definition (\ref{eq:decompose}) gives
\begin{equation}
	-\frac{1}{2}e^{-\phi}g^{(0)\alpha\mu}\frac{\partial}{\partial y}
		g^{(1)}_{\alpha\nu} = \Sigma^{(1)\mu}_{\quad\nu} + \frac{1}{4}
		\delta^{\mu}_{\nu} Q^{(1)} \ .
		\label{1:k}
\end{equation}	 
Integrating Eq.~(\ref{1:k}), we obtain the metric in the bulk:
\begin{eqnarray}
  	g^{(1)}_{\mu\nu} &=& -{l^2 \over 2 }\left({1\over a^2} -1 \right) 
    		( R_{\mu\nu}  - {1\over 6} h_{\mu\nu} R ) 
    		+ {l^2 \over 2} 
    		\left( {1 \over a^2 }-1 
    		-{2y e^{\phi}\over l}{1\over a^2} \right) 
    		\left( \phi_{|\mu \nu}  + {1\over 2} h_{\mu\nu}
   		\phi^{|\alpha}  \phi_{|\alpha} \right)   \nonumber \\
  		&& \quad   -{y^2 e^{2\phi} \over a^2} 
  		\left(  \phi_{|\mu}  \phi_{|\nu}-  {1\over 2} h_{\mu\nu}
   		\phi^{|\alpha}  \phi_{|\alpha}  \right)
    		-{l \over 2}\left({1\over a^4} -1 \right)   \chi_{\mu\nu} \ ,
    		\label{1:metric}
\end{eqnarray}
where we have imposed the boundary condition, 
$g^{(1)}_{\mu\nu} (y=0, x^\mu ) =0 $. 
 From these results, one can calculate the Weyl tensor as
\begin{equation}
   	C_{y\mu y\nu} = {2\chi_{\mu\nu} \over l a^4}  \ .
   	\label{eq:weyl}
\end{equation}
Hence, the term of $\chi_{\mu\nu}$ is essentially the Weyl tensor at 
this order. Note that we have obtained the bulk metric in terms of
$ \phi(x)$, $h_{\mu\nu}(x) $ and $ \chi_{\mu\nu}(x) $. 

\subsection{Quasi-Scalar-Tensor Gravity}

 We shall deduce the equations for $ \phi(x),\ h_{\mu\nu}(x) $ and 
 $ \chi_{\mu\nu}(x) $ from  junction conditions. 
 Using Eqs.~(\ref{1:Q}) and (\ref{1:sigma}),  
 one gets the junction conditions. 
The junction condition at the $A$-brane is written as
\begin{equation}
   	{l\over 2 } G^\mu_{\ \nu} (h ) + \chi^\mu_{\ \nu}
  		= {\kappa^2 \over 2} T^{A\mu}_{\quad \ \nu} \ .
  		\label{1:einstein-A}
\end{equation}
This equation is nothing but the Einstein equations with 
the  generalized dark radiation $\chi_{\mu\nu}$.
It should be noted that $\chi_{\mu\nu}$ is undetermined at this level,
exhibiting the non-local nature of Eq.~(\ref{1:einstein-A}). 

 The junction condition at 
the $B$-brane is given by
\begin{equation}
   	{l\over 2 \Omega^2 }  G^\mu_{\ \nu}   
   		+ { l e^{\phi} \over \Omega^2} \left(  \phi^{|\mu}_{\ |\nu} 
  		-\delta^\mu_\nu  \phi^{|\alpha}_{\ |\alpha} 
  		+  \phi^{|\mu}  \phi_{|\nu} -  \delta^\mu_\nu
   		\phi^{|\alpha}  \phi_{|\alpha} \right)  
   		+ {l e^{2\phi} \over \Omega^2 } \left( \phi^{|\mu} \phi_{|\nu}
  		+ {1\over 2} \delta^\mu_\nu  \phi^{|\alpha} \phi_{|\alpha} 
  		\right)
	    	+ {\chi^\mu_{\ \nu} \over \Omega^4}
  		= -{\kappa^2 \over 2 \Omega^2 } T^{B\mu}_{\quad \nu}  \ ,
  		\label{1:equation-B}
\end{equation}
where $\Omega (x) = a(y=l,x)= \exp[-e^{\phi}]$.
Here, the index of $T^{B\mu}_{\quad \ \nu}$ is the energy momentum tensor 
with the index raised by the induced metric $h_{\mu\nu}$ on the $A$-brane,
while $\tilde{T}^{B\mu}_{\quad \ \nu}$ is the one raised by the 
induced metric on the $B$-brane. At the present order, we
have the following relations,
\begin{eqnarray}
	T^{B}_{\mu\nu} = \tilde{T}^{B}_{\mu\nu} \ , \ 
	T^{B\mu}_{\quad\nu} = \Omega^{2}\tilde{T}^{B\mu}_{\quad\nu} \ .
	\label{rel:EM}
\end{eqnarray}
To reveal the role of the radion field, we must write  
 Eq.~(\ref{1:equation-B}) using the induced metric on the $B$-brane 
$g^{B{\rm brane}}_{\mu\nu}= \Omega^2 (h_{\mu\nu}+g^{(1)}_{\mu\nu}) 
\equiv f_{\mu\nu}+ \Omega^2g^{(1)}_{\mu\nu}$. 
At this order, the Ricci tensor $R^{\mu}_{\nu}$ of the induced metric
on the $B$-brane is equal to that of $f_{\mu\nu}$.   
Using this fact, we rewrite Eq.~(\ref{1:equation-B}) to obtain 
the effective equations on the $B$-brane,
\begin{equation}
 	{l\over 2 } G^\mu_{\ \nu} (f )  + {\chi^\mu_{\ \nu} \over \Omega^4}
 	=- {\kappa^2 \over 2 } \tilde{T}^{B\mu}_{\quad \nu}  \ .
 	\label{1:einstein-B}
\end{equation}
Again, we have the non-conventional (non-local) Einstein equations as 
in the case of the $A$-brane. 

Although Eqs.~(\ref{1:einstein-A}) and (\ref{1:einstein-B})
are non-local individually, with undetermined $\chi_{\mu\nu}$,
one can combine both equations to reduce them to local equations
for each brane. This happens to be possible since
$\chi_{\mu\nu}$ appears only algebraically; one can easily eliminate 
$\chi_{\mu\nu}$ from Eqs.~(\ref{1:einstein-A}) and (\ref{1:equation-B}). 
Defining a new field $\Psi = 1-\Omega^2$,  we find 
\begin{equation}
 	G^\mu_{\ \nu} (h) ={\kappa^2 \over l \Psi } T^{A\mu}_{\quad\ \nu}
      		+{\kappa^2 (1-\Psi ) \over l\Psi } T^{B\mu}_{\quad\ \nu}
      		+{ 1 \over \Psi } \left(  \Psi^{|\mu}_{\ |\nu} 
  		-\delta^\mu_\nu  \Psi^{|\alpha}_{\ |\alpha} \right)
  		+{\omega(\Psi ) \over \Psi^2} \left( \Psi^{|\mu}  \Psi_{|\nu}
  		- {1\over 2} \delta^\mu_\nu  \Psi^{|\alpha} \Psi_{|\alpha} 
  		\right)  \ ,
  		\label{1:STG-1}
\end{equation}
where the coupling function $\omega (\Psi)$ takes the following form:
\begin{equation}
  	\omega (\Psi ) = {3\over 2} {\Psi \over 1-\Psi }  \ .
  	\label{eq:coupling}
\end{equation}
We can also determine $\chi^{\mu}_{\nu}$ by eliminating $G^{\mu}_{\nu}$ 
from Eqs.~(\ref{1:einstein-A}) and (\ref{1:equation-B}). Then,
\begin{equation}
	\chi^{\mu}_{\ \nu} = -{\kappa^2 (1-\Psi) \over 2 \Psi} 
      		\left( T^{A\mu}_{\quad \nu} + T^{B\mu}_{\quad \nu} \right) 
      		-{l  \over 2 \Psi} \left[ \left(  \Psi^{|\mu}_{\ |\nu} 
  		-\delta^\mu_\nu  \Psi^{|\alpha}_{\ |\alpha} \right)
  		+{\omega(\Psi ) \over \Psi} \left( \Psi^{|\mu}  \Psi_{|\nu}
  		- {1\over 2} \delta^\mu_\nu  \Psi^{|\alpha} \Psi_{|\alpha} 
  		\right) \right]     \ .
  		\label{eq:chi}
\end{equation}
The condition $\chi^\mu_{\ \mu} =0$ gives rise to the field equation for 
$\Psi$:
\begin{equation}
  	\Box \Psi = {\kappa^2 \over l} {T^A + T^B \over 2\omega +3}
  		-{1 \over 2\omega +3}{d\omega \over d\Psi} \Psi^{|\mu} 
  		\Psi_{|\mu} \ ,
  		\label{1:STG-2}
\end{equation}
where we have used the explicit form of $\omega (\Psi) $. 
This equation tells us that the trace part of 
the energy momentum tensor determines the radion field and hence the relative
bending of the brane. 
And $\chi_{\mu\nu}$ is determined by the traceless part of the right-hand side 
of Eq.~(\ref{eq:chi}). Remarkably, $\chi_{\mu\nu}$ is now a secondary entity.

Eqs.~(\ref{1:STG-1}) and (\ref{1:STG-2}) are the basic equations 
to be used in cosmological or astrophysical contexts when the
characteristic energy density is less than $|\sigma_i|$.
 Notice that the conservation law with respect to the metric $h_{\mu\nu}$ 
reads 
\begin{equation}
   	T^{A\mu}_{\quad\ \nu |\mu } =0 \ , \quad
   	T^{B\mu}_{\quad\ \nu |\mu } = {\Psi_{|\mu} \over 1-\Psi }
        T^{B\mu}_{\quad\ \nu} 
        -{1\over 2}{\Psi_{|\nu} \over 1-\Psi}T^{B} \ .
        \label{eq:conserve-EM}
\end{equation}
In contrast to the usual scalar-tensor gravity, this theory  
couples with two kinds of matter, namely, the matters on both  positive 
 and  negative tension branes,  with different 
effective gravitational coupling constants. 
For this reason, we call this theory the quasi-scalar-tensor gravity.    
Thus, the (non-local) Einstein equations (\ref{1:einstein-A}) with 
the generalized dark radiation has transformed into 
the (local) quasi-scalar-tensor gravity (\ref{1:STG-1}) 
with the coupling function $\omega (\Psi) $. 

\subsection{Effective Action}

Let us consider an effective action for $h_{\mu\nu}(x)$ and $\phi(x)$. 
If one wants to calculate the quantum fluctuations in the inflationary 
scenario, for example, one needs the action to determine
 the magnitudes of them. 
The action have to be derived from the original 5-dimensional action by 
substituting the solution of the equations of motion in the bulk and 
integrating out over the bulk coordinate. 
We shall start with  the following action:
\begin{eqnarray}
  	S &=& {1\over 2\kappa^2 } \int d^5x \sqrt{-g}
  		\left[ {\cal R} + {12 \over l^2} \right]
  		+ {2\over \kappa^2} \int d^4 x \sqrt{-h} Q^A 
  		- {2\over \kappa^2} \int d^4 x \sqrt{-f} Q^B  \nonumber \\
  		&&   - {6\over \kappa^2 l} \int d^4 x \sqrt{-h}  
      		+ {6\over \kappa^2 l} \int d^4 x \sqrt{-f}  
      		+ \int d^4 x \sqrt{-h} {\cal L}^A 
      		+ \int d^4 x \sqrt{-f} {\cal L}^B       \ , 
      		\label{action:5-dim}
\end{eqnarray}
where we have taken into account the boundary term, the so-called 
Gibbons-Hawking term instead of introducing delta-function singularities
in the curvature. The factor 2 in the Gibbons-Hawking term comes from 
the $Z_2$ symmetry of this space-time. As we substitute the solution 
of the bulk equations of motion, we can use the 
the equation ${\cal R} = -20/l^2$ which holds in the bulk. 
 It should be stressed that the bulk metric is solved without using 
  junction conditions and is expressed in terms of $\phi$, $h_{\mu\nu}$ and 
  $\chi_{\mu\nu}$. That is why we can get the effective action on the brane
  by the simple substitution of the solution. 
 Now, up to the first order, we obtain
\begin{eqnarray}
  	S &=& -{8 \over \kappa^2 l^2 } \int d^4 x \sqrt{-h} \int_0^{le^\phi}
        	dz a^4 \left[ 1+ {1\over 2} h^{\mu\nu} g^{(1)}_{\mu\nu} \right]
        	+ {2\over \kappa^2 } \int d^4 x \sqrt{-h} \left[ {4\over l}
        	+ {l\over 6} R \right]  \nonumber\\
    		&& \quad    -{2\over \kappa^2} \int d^4 x \sqrt{-h} \Omega^4 
    		\left[
        	1+ {1\over 2} h^{\mu\nu} g^{(1)}_{\mu\nu} \right] 
        	\left[ {4\over l} +{l \over 6\Omega^2 } R +{le^\phi \over 
        	\Omega^2 }
        	\left( \Box \phi + \phi^{|\alpha} \phi_{|\alpha} \right)
        	- {le^{2\phi} \over \Omega^2 }\phi^{|\alpha} \phi_{|\alpha} 
        	\right] 
        	\nonumber \\ 
    		&&  \quad  -{6\over \kappa^2 l} \int d^4 x \sqrt{-h} 
        	+ {6\over \kappa^2 l} \int d^4 x \sqrt{-h} \Omega^4 \left[
        	1+ {1\over 2} h^{\mu\nu} g^{(1)}_{\mu\nu} \right] 
        	+ \int d^4 x \sqrt{-h} {\cal L}^A 
      		+ \int d^4 x \sqrt{-h} \Omega^4 {\cal L}^B        \ .
      		\label{pre-action:4-dim}
\end{eqnarray}
Using Eq.~(\ref{1:metric}) and the definition $\Psi = 1-\Omega^2 $, 
we finally have the action:
\begin{equation}
 	S  = {l \over 2 \kappa^2} \int d^4 x \sqrt{-h} 
     		\left[ \Psi R - {\omega (\Psi ) \over \Psi} 
     		\Psi^{|\alpha} \Psi_{|\alpha} \right]  
     		+ \int d^4 x \sqrt{-h} {\cal L}^A 
      		+ \int d^4 x \sqrt{-h} \left(1-\Psi \right)^2 {\cal L}^B  \ .  
      		\label{action:4-dim} 
\end{equation}
This is a complete derivation of the action with the correct 
normalization which is important for quantization of the theory. 
 
Here, it should be noted that $\chi^{\mu}_{\nu}$ which appeared
in $g^{(1)}_{\mu\nu}$ is a non-local quantity.
 In fact, eliminating $\Psi$ from Eq.~(\ref{eq:chi}) by solving 
Eq.~(\ref{1:STG-2}) yields a non-local expression of $\chi^{\mu}_{\nu}$. 
If we substitute this non-local expression into Eq.~(\ref{1:einstein-A}), 
we will obtain a non-local theory.  Conversely, one can see that
introducing the radion disentangles the non-locality in the 
non-conventional Einstein equations (\ref{1:einstein-A}) and yields the 
quasi-scalar-tensor
gravity given by Eqs.~(\ref{1:STG-1}) and (\ref{1:STG-2}).
 This important point is more transparent in the derivation of the 
 effective action. Indeed, the non-local part $\chi^{\mu}_{\nu}$ disappears 
 because of the traceless nature $\chi^{\mu}_{\mu}=0$.
 A similar mechanism is discussed 
 by Gen and Sasaki~\cite{gen} in the context of the linear theory.

\subsection{Holographic Brane Gravity}

We have obtained the 4-dimensional quasi-scalar-tensor gravity 
from the 5-dimensional action. 
The bulk metric corresponding to the 4-dimensional effective 
theory is  given by
\begin{equation}
  	g_{\mu\nu} = (1-\Psi )^{y/l} \left[
       		h_{\mu\nu} (x) + g^{(1)}_{\mu\nu} (h_{\mu\nu} , \Psi, 
        	T^A_{\mu \nu } ,  T^B_{\mu \nu } , y)  \right] \ , 
        	\label{holograms}
\end{equation}
where $\chi_{\mu\nu}$ in Eq.~(\ref{1:metric}) is eliminated 
by using Eq.~(\ref{eq:chi}). 
Here, the $y$-dependence of $g^{(1)}_{\mu\nu}$ is explicitly known. 
Thus the bulk metric is completely determined by the 
energy momentum tensors on both branes and the radion and the induced 
metric on the $A$-brane. Therefore, once the 4-dimensional solution of 
the quasi-scalar-tensor gravity is given, one can reconstruct the bulk 
geometry from these data.  The quasi-scalar-tensor gravity works as  
holograms at the low energy.  In this sense, one can call the 
quasi-scalar-tensor gravity the holographic brane gravity. 
Eq.~(\ref{holograms}) gives a holographic picture of 
the brane world. Recalling that the radion specifies the 
position of the second brane, the radion can be interpreted as a kind 
of ``phase" in the holographic picture  of the brane world.  

\subsection{Effective Theory on $B$-brane }
 
For completeness, we shall derive the effective equations of motion on 
the $B$-brane.  To do so, let us simply reverse the role of the 
$A$-brane and that of the $B$-brane.   
Substituting $h_{\mu\nu} =\Omega^{-2} f_{\mu\nu}$ 
into the junction conditions yields 
\begin{equation}
   	{l\over 2} G^\mu_{\ \nu} (f) + {\chi^\mu_{\ \nu} \over \Omega^4}
  		= -{\kappa^2 \over 2} T^{B\mu}_{\ \nu} 
\end{equation}
and
\begin{equation}
   	{l \Omega^2 \over 2}  G^\mu_{\ \nu}   
  		+ l \Omega^2  \left(  (\log\Omega)^{;\mu}_{\ ;\nu} 
  		-\delta^\mu_\nu  (\log\Omega)^{;\alpha}_{;\alpha} 
  		+  (\log\Omega)^{;\mu}  (\log\Omega)_{;\nu} 
  		+ {1\over 2}  \delta^\mu_\nu
   		(\log\Omega)^{;\alpha}  (\log\Omega)_{;\alpha} \right)  
    		+ \chi^\mu_{\ \nu} 
  		= {\kappa^2 \Omega^2 \over 2  } T^{A\mu}_{\ \nu}  \ ,
\end{equation}
where $;$ denotes the covariant derivative with respect to the metric 
$f_{\mu\nu}$. Thus, defining $\Phi = \Omega^{-2} -1$, 
we obtain the effective equation on the $B$-brane:
\begin{equation}
 	G^\mu_{\ \nu} (f) ={\kappa^2 \over l \Phi } T^{B\mu}_{\quad\ \nu}
      		+{\kappa^2 (1+\Phi ) \over l\Phi } T^{A\mu}_{\quad\ \nu}
      		+{ 1 \over \Phi } \left(  \Phi^{;\mu}_{\ ;\nu} 
  		-\delta^\mu_\nu  \Phi^{;\alpha}_{\ ;\alpha} \right)
  		+{\omega(\Phi ) \over \Phi^2} \left( \Phi^{;\mu}  \Phi_{;\nu}
  		- {1\over 2} \delta^\mu_\nu  \Phi^{;\alpha} \Phi_{;\alpha} 
  		\right)  \ ,
\end{equation}
where 
\begin{equation}
  	\omega (\Phi ) = -{3\over 2} {\Phi \over 1+ \Phi }  \ .
\end{equation}
The equations of motion for the radion becomes
\begin{equation}
  	\Box \Phi = {\kappa^2 \over l} {T^A + T^B \over 2\omega +3}
  		-{1 \over 2\omega +3}{d\omega \over d\Phi} \Phi^{;\mu} 
  		\Phi_{;\mu} \ .
\end{equation}
Thus, we have  shown that the gravity on the negative tension brane is 
described by  the quasi-scalar-tensor gravity with a coupling 
function  $\omega(\Phi) = -3\Phi /2(1+\Phi ) $. 

It should be noted that the dynamics on both  branes are not independent. 
We  know the gravity on the $B$-brane once we know that on the $A$-brane, 
and vice versa. The transformation rules are
\begin{eqnarray}
  	\Phi &=& {\Psi \over 1-\Psi}  \ , \\
  		g^{B{\rm brane}}_{\mu\nu} &=& (1-\Psi ) \left[ h_{\mu\nu} 
        	+ g^{(1)}_{\mu\nu} 
        	\left(     h_{\mu\nu} , \Psi, 
        	T^A_{\mu\nu } ,  T^B_{\mu\nu } , \ 
        	y=l \right) \right] \ .
\end{eqnarray}
This relation is useful when we consider concrete applications.

\section{Kaluza-Klein corrections}

As explained in Appendix A, our analysis so far to the first order in
the gradient expansion corresponds to the zero-mode truncation
in the language of the linearized theory. 
Although it is obscure to use the word ``Kaluza-Klein corrections" in 
the non-linear theory, we shall call their non-linear counterpart 
simply as Kaluza-Klein corrections in this paper. 

In principle, we can continue our analysis up to a desired order 
using the following recursive formulas: 
\begin{eqnarray}
    	& & \Sigma^{(n) \mu}_{\quad \   \nu }  
    		= - {1\over a^4} \int dy a^4 \left\{
    		\left[ \R - \nabla^\mu \nabla_\nu \phi 
      		- \nabla^\mu \phi \nabla_\nu \phi \right]^{(n)}_{\rm traceless 
      		\ part}
        	-\sum_{p=1}^{n-1} Q^{(p)} 
        	\Sigma^{(n-p) \mu}_{\quad \quad  \nu} \right\} \ , \\
   	&  & Q^{(n)} = {l\over 6} \sum_{p=1}^{n-1}
        	\left[-{3\over 4} Q^{(p)}  Q^{(n-p)}
      		+  \Sigma^{(p)\alpha}_{\quad \  \beta} 
      		\Sigma^{(n-p)\beta}_{\quad \quad  \alpha} 
      		+ [\Rs]^{(n)}  \right] , \\
   	& & Q^{(n)}_{, y} -{2\over l}Q^{(n)} 
   		=  \sum_{p=1}^{n-1} \left\{ {1\over 4} Q^{(p)} Q^{(n-p)}
     		+ \Sigma^{(p)\alpha \beta} \Sigma^{(n-p)}_{\alpha \beta} 
        	\right\}  + \left[ 
        	\nabla^\alpha \nabla_\alpha \phi 
       		+ \nabla^\alpha \phi \nabla_\alpha \phi \right]^{(n)}
        	\ , \\
   	& & \Sigma_{\mu \quad \ |\lambda}^{(n)\ \lambda}  
    		-  {3\over 4} Q^{(n)}_{|\mu} 
    		+ \sum_{p=1}^{n-1} \left\{ \Gamma^{(p)\alpha }_{\lambda \alpha}
    		\Sigma^{(n-p)\lambda}_{\quad\quad \mu}
       		- \Gamma^{(p)\lambda }_{\alpha\mu} 
       		\Sigma^{(n-p)\alpha}_{\quad\quad \lambda} \right\}=0 \ . 
\end{eqnarray}
These equations give a solution as an infinite sum. 
The existence of the infinite series is a manifestation of the non-locality 
of the brane model~\cite{mukoh}. 
 
To get the effective equations of motion with second 
order corrections using the above formula  is straightforward. 
However, carrying out the calculation is laborious 
and the resultant expression is too long to write down. 
 As for the linear theory, we will obtain the explicit effective equations of
  motion with Kaluza-Klein corrections in Appendix B. 
 Here, we will only sketch how the Kaluza-Klein corrections appear using an 
 easy method. 

Although we need the explicit $y$-dependence of the bulk to obtain the
action, as long as we are interested only in the effective equations
on the brane, we do not have to solve the bulk explicitly.
The reason is as follows. We can write down the non-local Einstein 
equations corresponding to 
 Eqs.~(\ref{1:einstein-A}) and  (\ref{1:einstein-B}) without 
knowing the bulk geometry. Then since we know how the non-local term, i.e.,
the generalized dark radiation term behaves in the bulk, we may eliminate
it just as in the 1-st order case.

The non-local Einstein equations on the branes are~\cite{kanno} 
\begin{eqnarray}
  	G^{(4)}_{\mu\nu} (h ) &=& 
  		-{2\over l}\left( \chi_{\mu\nu} + t_{\mu\nu} \right)
  		+{\kappa^2 \over l} T^A_{\mu\nu}  
   		-{l^2 \over 2} {\cal S}_{\mu\nu} -{l^2 \over 12} 
   		\left( RR_{\mu\nu} - {1\over 2} h_{\mu\nu} R^2 
   		+{3\over 4} h_{\mu\nu} R^\alpha_\beta R^\beta_\alpha \right) 
   		\nonumber\\
		&&  +{l \over 2}\left[ \chi^\alpha_{\ \mu |\nu\alpha} 
  		+\chi^\alpha_{\ \nu |\mu\alpha} -\chi_{\mu\nu |\alpha}^{\quad |
  		\alpha} 
  		\right]  - l  \chi_{\mu\alpha} R^\alpha_{\ \nu}    
   		+ {l \over 6} R\chi_{\mu\nu}  
   		-{1\over 4}h_{\mu\nu} \chi^\alpha_\beta \chi^\beta_\alpha  
   		\label{2:einstein-A} \\ 
  	G^{(4)}_{\mu\nu} (g^{B{\rm brane}} ) &=& 
  		-{2\over l\Omega^4 }\left( \chi_{\mu\nu} + t_{\mu\nu} \right) 
    		- {\kappa^2 \over l} T^B_{\mu\nu}  
   		-l^2\left( 1+ {\Omega^2 \over 2} \right){\cal S}_{\mu\nu}  
   		-{l^2 \over 12} \Omega^2 \left( RR_{\mu\nu} 
   		- {3\over 2} g^{B{\rm brane}}_{\mu\nu } 
   		R^\alpha_\beta R^\beta_\alpha  
   		\right) \nonumber \\
 		&& +{l \over 4}\left( 1+ {1 \over \Omega^4 } \right) 
 		\left[ \chi^\alpha_{\ \mu ;\nu\alpha}
  		+ \chi^\alpha_{\ \nu ;\mu\alpha} 
  		-\chi_{\mu\nu ;\alpha}^{\quad ;\alpha} 
  		\right]  -{l \over 2}\left( 1+ {1 \over \Omega^4 } \right) 
  		\left( \chi_{\mu\alpha} R^\alpha_{\ \nu} -{1\over 4} 
  		g^{B{\rm brane}}_{\mu\nu }
  		\chi_{\alpha\beta} R^{\alpha \beta} \right)   
   		+ {l \over 6\Omega^4} R\chi_{\mu\nu}     \nonumber \\
  		&& -{3\over 4} g^{B{\rm brane}}_{\mu\nu} \left[ 
  		{l^2\over 4} \left( R^\alpha_\beta R^\beta_\alpha -{2\over 9} 
  		R^2 \right) + {l \over 2 \Omega^4 } \chi^\alpha_\beta 
  		R^\beta_\alpha
  		-{l \over 6}\chi^\alpha_\beta R^\beta_\alpha
  		+{1\over 3 \Omega^8 } \chi^\alpha_\beta \chi^\beta_\alpha 
  		\right]  \ . 
  		\label{2:einstein-B}
\end{eqnarray}
where $t_{\mu\nu}$ is an integration constant at the 2-nd order
 and we have defined the quantity
\begin{eqnarray}
  	{\cal S}^\mu_{\ \nu} &=& R^\mu_{\ \alpha} R^\alpha_{\ \nu}
     	-{1\over 3} R R^\mu_{\ \nu} 
        -{1\over 4} \delta^\mu_\nu (R^\alpha_{\ \beta} R^\beta_{\ \alpha}
        - {1\over 3} R^2)  \nonumber \\ 
    	& & \qquad -{1\over 2} \left( R^{\alpha\mu}_{\ \ |\nu\alpha}
        + R^{\alpha \ |\mu}_{\ \nu \ \  | \alpha}  
        -{2\over 3} R^{|\mu}_{\ |\nu}  - \Box R^\mu_{\ \nu} 
        +{1\over 6} \delta^\mu_\nu \Box R \right)   \ .
        \label{eq:s}
\end{eqnarray}
Here, $;$ represents the covariant derivative with respect to $f_{\mu\nu}$ 
and all of the curvatures in Eq.~(\ref{2:einstein-A}) are calculated 
from $g^{B{\rm brane}}_{\mu\nu}$. 
What we should do is to eliminate $t_{\mu\nu}$ from Eqs.~(\ref{2:einstein-A}) 
and (\ref{2:einstein-B}) and 
substitute the relation 
$g^{B{\rm brane}}_{\mu\nu} = \Omega^2 [ h_{\mu\nu} + g^{(1)}_{\mu\nu}] $ 
into the resulting equation. Then we obtain a higher derivative but local
theory on the brane. 

Noticeably, the same is true for all higher order corrections. 
Thus, one can infer that 
the radion disentangles the non-locality in the system 
to all orders at the expense of introducing higher derivative terms.

\section{Conclusion}

We have developed a method to deduce the low energy effective theory 
for the two-brane system. 
The 5-dimensional equations of motion in the bulk is solved 
using a low energy expansion method. 
This allows us, through the junction conditions, 
to deduce the effective equations of motion for the gravity on the brane. 
As a result, we  have shown that the gravity 
on the brane world is described by a quasi-scalar-tensor theory with 
a specific coupling function 
$\omega(\Psi) = 3\Psi /2(1-\Psi ) $ on the positive tension brane
and $\omega(\Phi) = -3\Phi /2(1+\Phi ) $ on the negative tension brane,
where $\Psi$ and $\Phi$ are Brans-Dicke-like scalars on the positive
and negative tension branes, respectively. 
In contrast to the usual scalar-tensor theory, the quasi-scalar-tensor 
theory couples with matters on both branes but with different 
effective gravitational coupling constants.  In particular, 
the radion disguised as the scalar fields $\Psi$ and $\Phi$ couples  with 
the sum of the traces of the energy  momentum tensor on both branes.  
Moreover, we have derived the effective action by substituting the 
solution of the bulk equations of motion into the original action. 
This direct method determines the normalization of the effective action 
which is indispensable for quantizing the theory. 

In the process of derivation of the effective equations of motion,  we have 
clarified how the quasi-scalar tensor gravity emerges from  
Einstein's theory with the generalized dark radiation term
described by $\chi_{\mu\nu}$.   
A brane can feel the non-local effect  of the bulk geometry only through 
 $\chi_{\mu\nu}$ irrespective of the 
existence of another brane.  This is the picture that the Einstein 
equations with the generalized dark radiation tells us. 
Then, what is the role of the  radion? 
In order to make the connection between the radion and $\chi_{\mu\nu}$, 
we have to know the bulk geometry. 
In the case of a single brane, 
$\chi_{\mu\nu}$ is determined by the boundary conditions at the 
Cauchy horizon. If we require that the geometry is asymptotically 
 Anti-deSitter there, then $\chi_{\mu\nu}$ must 
vanish~\cite{wiseman2}. 
In the two-brane case, we have no asymptotic region, instead we have the 
second brane in the bulk. The radion determines
the location of the second brane where the junction conditions are imposed. 
The junction conditions give  $\chi_{\mu\nu}$  as a function of  
the energy momentum  tensor and the radion.   
The resultant equation is nothing but the holographic 
quasi-scalar-tensor gravity.  Thus, the difference 
between the Einstein equations with the generalized dark radiation and 
the quasi-scalar-tensor gravity is just superficial.  
The radion has converted the non-local non-conventional 
Einstein equations to the local quasi-scalar-tensor gravity. 

We have also given a systematic method to calculate the corrections 
due to Kaluza-Klein massive modes. 
It is conjectured that all of the non-locality arising from the integration
 is disentangled by the radion in the two brane system. 
We have also emphasized the holographic aspect of our result. 
It turns out that the effect of the bulk gravity on the low energy 
physics in the brane world can be described solely in the 4-dimensional 
language.  Conversely, 
the bulk geometry can be reconstructed from the knowledge of 
the 4-dimensional data. In this sense,  the quasi-scalar-tensor 
gravity we have found in this paper works as  holograms and hence 
can be called the holographic brane gravity.  

Let us discuss an implication of our results.
Cosmology is usually formulated on the basis of local 
field  theory. 
However, the superstring theory suggests  non-local field theories are 
ubiquitous. 
Though a non-local field theory is not easy to treat properly, the 
holographic description opens a new possibility to study cosmology with 
non-local terms. The brane world cosmology can be regarded as a 
realization of a non-local field theoretic approach to cosmology. 
In the single brane picture, the non-local terms due to the integration
 constant appear~\cite{kanno}. 
Furthermore, there are infinite series of higher derivative terms 
in the low energy expansion scheme.  This is also a manifestation of the 
non-locality of the brane world gravity~\cite{kanno,mukoh}. 
In the two brane system,  there also exist 
the above two types of non-locality. 
Intriguingly, the radion has disentangled the non-locality 
of the homogeneous solutions and led to 
the quasi-scalar-tensor gravity. 
Hence,  the quasi-scalar tensor theory is a 
non-local theory disguised as a local theory. 
In fact, integrating out the scalar field yields a non-local field 
theory.  In addition, the non-locality due to the Kaluza-Klein type  
corrections remains as an infinite series in the low energy expansion 
even in the two brane system.  Cosmology with non-local fields 
from this point of view deserves  further investigation. 

As we have succeeded to obtain the effective action for the non-linear brane 
gravity,  various problems can be now investigated. The two brane inflation 
is  under investigation using our method~\cite{sugumi}. 
Astrophysical applications 
such as gravitational waves from binary stars are also intriguing.  
Extension of our formalism to more general models which include bulk scalars
 or vector fields might be interesting.

\begin{acknowledgements}
We would like to thank M. Sasaki for valuable suggestions which improved the 
 presentation of the paper significantly. 
This work was supported in part by  Monbukagakusho Grant-in-Aid No.14540258. 
\end{acknowledgements}

\appendix

\section{Scalar field Example} 

In order to illustrate the method used in the main text, we examine 
a toy model in this appendix. Let us consider a massless scalar field 
$\phi$ in the background 
\begin{equation}
  	ds^2 = dy^2 + \exp[-2{y\over l} ] \eta_{\mu\nu} dx^\mu dx^\nu \ ,
  	\label{metric:background}
\end{equation}
where the branes are located at $y=0$ and $y=d$. 
The equation of motion for $\phi$ with a source on the branes 
becomes
\begin{equation}
  	\Box^{(5)} \phi 
  		= e^{4 y/l}\partial_y \left[ e^{-4y/l} \partial_y \phi \right]
   		+ e^{2y/l} \Box \phi = J^{A}(x) \delta (y) 
        	+ J^{B} (x) \delta (y-d)    \ .
        	\label{eq:5-dim}
\end{equation}
 From this equation, one can deduce 
the junction conditions 
\begin{eqnarray}
  	\partial_y \phi |_{y=0} &=& {1\over 2} J^{A} (x)  \ ,
  	\label{JC-s:A} \\ 
  	\partial_y \phi |_{y=d} &=& -{1\over 2} J^{B} (x) \ .
  	\label{JC-s:B}
\end{eqnarray}
Let us focus on the $A$-brane at $y=0$ and put 
\begin{equation}
   	\phi (y=0 , x) = \phi_0 (x)  \ .
   	\label{BC-s}
\end{equation}
The Green function with the Neumann boundary condition is easily calculated as 
\begin{eqnarray}
  	\Delta_5 (0, x;  0, x' ) &=& \int  {d^4 p \over (2\pi )^4}
  		\exp[ip \cdot (x-x') ] {1\over q}
  		{J_1 (ql e^{d/l} ) H^{(1)}_2 (ql) 
  		- J_2 (ql ) H^{(1)}_1 (ql e^{d/l} )  
  		\over J_1 (ql e^{d/l} ) H^{(1)}_1 (ql) 
  		- J_1 (ql) H^{(1)}_1 (ql e^{d/l})  }   \nonumber \\
 		&=& \int  {d^4 p \over (2\pi )^4}  \exp[ip \cdot (x-x') ] 
  		{2 \over q^2 l (1- e^{-2d/l})} \left[ 1 +
  		q^2 l^2 \left( {3\over 8} -{1\over 8} e^{-2d/l} 
  		-{1\over 2(1-e^{-2d/l}) } {d\over l} \right) +\cdots 
  		\right]  \ ,
  		\label{func:green-1}
\end{eqnarray}
and
\begin{eqnarray}
  	\Delta_5 (0, x;  d, x' ) &=& -{2i e^{d/l} \over \pi l }
        	 \int  {d^4 p \over (2\pi )^4}
        	\exp[ip \cdot (x-x') ] {1\over q^2 }
   		{1  \over  J_1 (ql e^{d/l} ) H^{(1)}_1 (ql) 
   		- J_1 (ql) H^{(1)}_1 (ql e^{d/l})  }  \nonumber \\
   		&=& \int  {d^4 p \over (2\pi )^4}  \exp[ip \cdot (x-x') ] 
  		{2 \over q^2 l (1- e^{-2d/l} )} \left[ 1 +
  		q^2 l^2 \left( {1\over 8} +{1\over 8} e^{2d/l} 
  		-{1\over 2(1-e^{-2d/l}) } {d\over l} \right) +\cdots 
  		\right]   \ ,
  		\label{func:green-2}
\end{eqnarray}
where $q^2 =-\eta_{\mu\nu}p^{\mu}p^{\nu} $. 
Thus, the standard Green function method gives the solution 
for Eq.~(\ref{eq:5-dim}) as
\begin{equation}
  	\phi_0 (x) = {1\over 2} \int d^4 x' \Delta_{5} (0,x; 0,x') 
  		J^A (x' ) + {1\over 2} \int d^4 x' e^{-4d/l} 
  		\Delta_{5}  (0,x; d,x') J^B (x' ) \ .
  		\label{solution:scalar}
\end{equation}
This gives 
\begin{eqnarray}
  	\Box \phi_0 (x) &=& {1\over 2} \int d^4 x' \Box \Delta_{5} (0,x; 0,x') 
  		J^A (x' ) + {1\over 2} \int d^4 x' e^{-4d/l} 
  		\Box \Delta_{5}  (0,x; d,x') J^B (x' ) \nonumber \\	
  		&=& {1\over l (1- e^{-2d/l})}  J^A (x )
    		+ {e^{-4d/l} \over l (1- e^{-2d/l})} e^{-4d/l} J^B (x ) 
    		\nonumber \\
 		&& \quad   + {1\over l (1- e^{-2d/l})}  \left[ 
  		{3\over 8} -{1\over 8} e^{-2d/l} 
  		-{1\over 2(1-e^{-2d/l}) } {d\over l} \right] l^2 \Box J^A (x) 
  		\nonumber \\
 		&& \quad +{e^{-4d/l} \over l (1- e^{-2d/l})} e^{-4d/l}
 		\left[
 		{1\over 8} +{1\over 8} e^{2d/l} 
 		-{1\over 2(1-e^{-2d/l}) } {d\over l} \right] l^2 \Box J^B (x) 
 		+ \cdots  \ .
 		\label{eq:4-dim}
\end{eqnarray}
Note that the first two terms come from the zero mode and the rest are  
Kaluza-Klein corrections. 
Now, we shall compare the above result, Eq.~(\ref{eq:4-dim}),
 with our method.

\subsection{0-th order}

At the 0-th order, we ignore gradients on the brane, then we get
\begin{equation}
  	e^{4 y/l}\partial_y \left[ e^{-4y/l} \partial_y \phi^{(0)} \right] =0 
  	\ .
  	\label{0:eq}
\end{equation}
The solution of Eq.~(\ref{0:eq}) is given by 
\begin{equation}
   	\phi^{(0)} = \phi_0 + e^{4 y/l} \psi_0 \ .
   	\label{0:phi}
\end{equation}
However, as we are regarding the source terms as the first order quantities, 
the junction conditions~(\ref{JC-s:A}) and (\ref{JC-s:B}) imply 
$\psi_0 =0 $. 
Hence, we simply obtain $\phi^{(0)} = \phi_0 $.

\subsection{1-st order}

At the first order, we must solve
\begin{equation}
   	e^{4 y/l}\partial_y \left[ e^{-4y/l} \partial_y \phi^{(1)} \right] =
   		- e^{2y/l} \Box \phi_0 (x) \ .
   		\label{1:eq}
\end{equation}
The result is
\begin{equation}
  	\phi^{(1)} = {l^2 \over 4} e^{2y/l} \Box \phi_0 
        	+ {l\over 4}e^{4y/l} C(x) + D(x) 
        	\label{A1:phi}
\end{equation}
where $C$ and $D$ are homogeneous solutions. 
The junction conditions~(\ref{JC-s:A}) and (\ref{JC-s:B}) 
become
\begin{eqnarray}
   	\partial_y \phi |_{y=0} &=& {l\over 2} \Box \phi_0 + C
   		= {1\over 2} J^{A} (x)    \ , 
   		\label{1:JC-s-A} \\
    	\partial_y \phi |_{y=d} &=& {l \over 2} e^{2d/l} \Box \phi_0 
    		+ C e^{4d/l}  = -{1\over 2} J^{B} (x)  \ .
    		\label{1:JC-s-B}
\end{eqnarray}
Eliminating $C$ from these equations,
we obtain 
\begin{equation}
  	\Box \phi_0 = {1\over l (1- e^{-2d/l})} J^A 
        	+ {e^{-4d/l} \over l (1- e^{-2d/l})}  J^B  \ .
        	\label{1:equation}
\end{equation}
This agrees with the zero-mode part of Eq.~(\ref{eq:4-dim}). 
Thus our method
to the first order corresponds to the zero-mode truncation when linearized. 
The homogeneous part is also determined as 
\begin{equation}
  	C(x) = {1\over 2 (1-e^{2d/l})}\left[ J^A +e^{-2d/l} J^B \right]  \ .
  	\label{eq:c}
\end{equation}

\subsection{2-nd order}

At the 2-nd order, we have
\begin{equation}
   	e^{4 y/l}\partial_y \left[ e^{-4y/l} \partial_y \phi^{(2)} \right] =
   		- e^{2y/l} \Box \phi^{(1)} (x) \ .
   		\label{2:eq}
\end{equation}
This Eq.~(\ref{2:eq}) can be integrated as
\begin{equation}
  	\partial_y \phi^{(2)} = -{l^2 \over 4} \left( ye^{4y/l} 
  		+{l\over 2} e^{2y/l} \right) \Box^2 \phi_0 
  		-{l^2 \over 8} (e^{6y/l} + e^{2y/l} ) \Box C(x)  
  		+ e^{4y/l} D(x)  \ .
  		\label{2:phi}  		
\end{equation}
Hence, the junction conditions~(\ref{JC-s:A}) and (\ref{JC-s:B}) yield
\begin{eqnarray}
  	&&{l\over 2} \Box \phi_0 + C(x)  -{l^3 \over 8} \Box^2 \phi_0 
  		-{l^2 \over 4} \Box C(x) + D(x) 
   		= {1\over 2} J^{A} (x)    \ , 
   		\label{2:JC-s-A} \\
 	&& {l \over 2} e^{2y/l} \Box \phi_0  + C(x) e^{4y/l}  
  		-{l^2 \over 4} (d e^{4y/l} + {l\over 2} e^{2y/l} ) 
  		\Box^2 \phi_0
  		-{l^2 \over 8} (e^{6y/l} + e^{2y/l} ) \Box C(x)
  		+ e^{4y/l} D(x) = -{1\over 2} J^{B} (x)  \ .
  		\label{2:JC-s-B}
\end{eqnarray}
Combining both Eqs.~(\ref{2:JC-s-A}) and (\ref{2:JC-s-B}), we get
\begin{equation}
  	\Box \phi_0 = {1\over l (1- e^{-2d/l})}
  	\left[ J^A + e^{-4d/l} J^B \right]
  	+\left[ {l^2 \over 4} -{ld\over 2} {1\over (1-e^{-2d/l})}
  	\right] \Box^2 \phi_0 -{l\over 4} e^{2d/l} 
  	\left[ 1-e^{-2d/l} \right] \Box C    \ .
  	\label{2:equation}
\end{equation}
Substituting Eqs.~(\ref{1:equation}) and (\ref{eq:c}) into 
the right-hand side of Eq.~(\ref{2:equation}) 
yields Eq.~(\ref{eq:4-dim}). 
Thus, we have shown that the 2-nd order equations in our method correspond
to taking into account
the Kaluza-Klein corrections when the equations are linearized.

\section{Linearized Gravity}

Let us now turn to the case of our interest, that is, the linearized gravity.
In the linearized gravity, following the method in \cite{GKR}, 
the solution is explicitly given 
in terms of the scalar Neumann Green function $\Delta_{5}$ in 
Eqs.~(\ref{func:green-1}) and (\ref{func:green-2});
\begin{eqnarray}
  	\bar{h}_{\mu\nu}^{A} (x^\mu ) &=& -\kappa^2 \int d^4 x' \Delta_{5} 
  	(0,x^{\mu}; 0,x^{\mu '}) \left[ T_{\mu\nu}^A (x' ) - 
 	\frac{1}{6}\eta_{\mu\nu} T^{A}(x') \right] \nonumber \\
  	& & - \frac{1}{3}\frac{\kappa^2}{l}\eta_{\mu\nu}
  	\int d^4 x' \Delta_{4}(x^{\mu}, x^{\mu '}) T^{A}(x') \nonumber \\
  	& & - \kappa^2 \int d^4 x' e^{-2d/l} \Delta_{5} 
  	(0,x^{\mu}; d,x^{\mu '})
  	\left[ T_{\mu\nu}^B (x' ) - 
  	\frac{1}{6}\eta_{\mu\nu} T^{B}(x') \right] \ ,
  	\label{eq:h}
\end{eqnarray}
where $\bar{h}_{\mu\nu}^{A}$ is the small fluctuation in the metric on
the $A$-brane. Applying $\Box$ to this equation and expanding $\Delta_{5}$
as Eqs.~(\ref{func:green-1}) and (\ref{func:green-2}), we obtain
\begin{eqnarray}
	\Box\bar{h}_{\mu\nu}^{A}
		&=& -\frac{2\kappa^2}{l}\frac{1}{1-e^{-2d/l}}
		\left( \stackrel{A}{T_{\mu\nu}} + e^{-2d/l}
		\stackrel{B}{T_{\mu\nu}} \right) +
		\frac{1}{3}\frac{\kappa^2}{l}\frac{e^{-2d/l}}{1-e^{-2d/l}}
		\eta^{\mu}_{\nu}
		\left( \stackrel{A}{T} + \stackrel{B}{T} \right) \nonumber \\
		& & -2\frac{\kappa^2}{l}\frac{1}{1-e^{-2d/l}}
		\left[ \frac{3}{8}-\frac{1}{8}e^{-2d/l} 
		- \frac{d/l}{2(1-e^{-2d/l})}
		\right] l^2\Box \left( \stackrel{A}{T_{\mu\nu}} - \frac{1}{6}
		\eta_{\mu\nu}\stackrel{A}{T} \right) \nonumber \\
		& & - 2\frac{\kappa^2}{l}\frac{e^{-2d/l}}{1-e^{-2d/l}}
		\left[ \frac{1}{8}+\frac{1}{8}e^{2d/l} - 
		\frac{d/l}{2(1-e^{-2d/l})}\right] 
		l^2\Box \left( \stackrel{B}{T_{\mu\nu}} - \frac{1}{6}
		\eta_{\mu\nu}\stackrel{B}{T} \right) \ .
		\label{eq:g-wave}
\end{eqnarray}
This may be regarded as the effective Einstein equations corrected to 
$O((\frac{l}{L})^4)$.
Now we demonstrate that our low-energy expansion scheme leads to the 
linearized quasi-Brans-Dicke gravity.
Then we will show that our method correctly reproduces Eq.~(\ref{eq:g-wave}). 

Our solution for the bulk metric is
\begin{equation}
 	ds^2 = e^{2\phi (y,x) } dy^2  +  
        	\exp\left[-{2\over l} \int dy e^{\phi (y,x)}\right] 
           	h_{\mu\nu}(x^\mu ) dx^\mu dx^\nu  \ .
           	\label{metric:bulk} 
\end{equation}
Here, two branes are located at $ y = 0$ and $y=l$. 
We will assume that $\phi(y,x^{\mu}) \equiv \phi(x^{\mu})$ for simplicity.
After some obvious  changes of variables and rescalings of coordinates,
 small fluctuations in the metric can be represented as
\begin{equation}
 	ds^2 = (1+2\delta \phi)dy^2  +  
        	e^{-\frac{2}{l}y} \left(\eta _{\mu\nu} + h_{\mu\nu}(x^\mu ) -
        	\frac{2y}{l}\eta_{\mu\nu}\delta\phi (x^\mu)
        	\right)dx^\mu dx^\nu  \ , 
        	\label{metric:fluctuation}
\end{equation}
thus 
\begin{equation}
	\stackrel{(0)}{\delta g_{\mu\nu}}(x,y) = 
		a^2\left[ h_{\mu\nu}(x) 
		- \frac{2y}{l}\eta_{\mu\nu}\delta\phi (x) 
		\right], \quad a=e^{-\frac{y}{l}} \ ,
		\label{0:metric}
\end{equation}
where $h_{\mu\nu}$ and $\delta \phi$ represent tensor and scalar 
fluctuations, respectively. Now the
two branes are located at $y=0$ and $y=d$, 
because of the relation $e^{\phi}=d/l$.  Decomposing the extrinsic 
curvature into the traceless part and the trace part,
the small fluctuations of each part are 
\begin{equation}
	\Sigma^{\mu}_{\nu} = \delta\Sigma^{\mu}_{\nu},\quad
	Q = \frac{4}{l} + \delta Q \ .
	\label{fluctuation:k}
\end{equation}
Here we used our results in Eq.~(\ref{0:Q}).
The equations off the brane,
 Eqs.~(\ref{eq:munu-traceless}-\ref{eq:ymu}), are linearized
to become
\begin{eqnarray}
 	& & \delta\Sigma^\mu_{\ \nu , y} 
 		- \frac{4}{l} \delta\Sigma^\mu_{\ \nu} 
    		= -\left[ R^{\mu}_{\ \nu} - {1\over 4} \delta^\mu_\nu R 
            	-\nabla^\mu \nabla_\nu \delta\phi 
               	+{1\over 4} \delta^\mu_\nu 
               	\nabla^\alpha \nabla_\alpha \delta\phi
               	\right]      \ ,
               	\label{linear:munu-traceless}     \\
 	& &  \delta Q 
     		= \frac{l}{6}\left[ R \right]  \ , 
     		\label{linear:munu-trace} \\
 	& &  \delta Q_{, y} -{2\over l}\delta Q 
     		= \nabla^\alpha \nabla_\alpha \delta\phi  \ ,
     		\label{linear:yy}  \\
 	& &  \nabla_\lambda \delta\Sigma_{\mu }^{\ \lambda}  
           	-  {3\over 4} \nabla_\mu \delta Q = 0   \ .
           	\label{linear:ymu}
\end{eqnarray}
The junction conditions become
\begin{eqnarray}
   	\left[ \delta\Sigma^{\mu}_{\ \nu} 
   		- \frac{3}{4} \delta^\mu_\nu \delta Q \right] 
   		\Bigg|_{y=0}
    		&=& \frac{\kappa^2}{2}\stackrel{A}{T^\mu_{\ \nu}}  \ , 
    		\label{JC:linear-A} \\
   	\left[ \delta\Sigma^{\mu}_{\ \nu} 
   		- \frac{3}{4} \delta^\mu_\nu \delta Q \right] 
   		\Bigg|_{y=d}
    		&=& -\frac{\kappa^2}{2}
    		\stackrel{B}{\tilde{T}^\mu_{\ \nu}} \ .
    		\label{JC:linear-B}
\end{eqnarray}
We now work with our low-energy iteration scheme. The goal is to construct the
metric fluctuation as 
\begin{equation}
	\delta g_{\mu\nu}(x,y) = 
		a^2\left[ h_{\mu\nu}(x) 
		- \frac{2y}{l}\eta_{\mu\nu}\delta\phi (x) +
		\stackrel{(1)}{\delta g_{\mu\nu}}(x,y) +
		\stackrel{(2)}{\delta g_{\mu\nu}}(x,y) + \cdots
		\right]. \quad 
		\label{fluctuation:g}
\end{equation}
\subsection{1-st order}

The solution at the 1-st order is 
\begin{eqnarray}
	\stackrel{(1)}{\delta Q} &=& \frac{l}{6a^2} R(h) +
		\frac{y}{a^2}\Box\delta\phi \ , 
		\label{1:linear-Q} \\	
	\stackrel{(1)}{\delta\Sigma^{\mu}_{\ \nu}} &=& \frac{l}{2a^2} 
		\left( R^{\mu}_{\ \nu} 
		- \frac{1}{4}\delta^{\mu}_{\nu}R \right) +
		\frac{y}{a^2}\left( \delta\phi^{|\mu}_{\ |\nu} - 
		\frac{1}{4}\delta^{\mu}_{\nu}\Box\delta\phi \right) + 
		\frac{\chi^{\mu}_{\ \nu}}{a^4},\quad \chi^{\mu}_{\ \mu}=0 \ ,
		\label{1:linear-sigma} \\
	\stackrel{(1)}{\delta K^{\mu}_{\ \nu}} &=& \frac{l}{2a^2} 
		\left(R^{\mu}_{\nu} - \frac{1}{6}\delta^{\mu}_{\ \nu} R \right)
		+ \frac{y}{a^2}\delta\phi^{|\mu}_{\ |\nu} + 
		\frac{\chi^{\mu}_{\ \nu}}{a^4} \ , 
		\label{1:linear-k} \\
	\stackrel{(1)}{\delta g_{\mu\nu}} &=& -\frac{l^2}{2} 
		\left(\frac{1}{a^2}-1 \right) 
		\left(R_{\mu\nu} - \frac{1}{6}\eta_{\mu\nu} R \right)
		- \left( \frac{ly}{a^2}-\frac{l^2}{2a^2}+\frac{l^2}{2} \right)
		\delta\phi_{|\mu\nu} -\frac{l}{2}\left(\frac{1}{a^4}-1 \right)
		\chi_{\mu\nu} \ .
		\label{1:linear-g}
\end{eqnarray}
 From Eq.~(\ref{linear:ymu}), we obtain the constraint 
$\chi^{\mu}_{\ \nu|\mu}=0$ for the 
homogeneous solution. The junction conditions are 
\begin{eqnarray}
   	\left[ \stackrel{(1)}{\delta\Sigma^{\mu}_{\ \nu}} 
   		- \frac{3}{4} \delta^\mu_\nu \stackrel{(1)}{\delta Q} \right] 
   		\Bigg|_{y=0}
    		&=& \frac{l}{2}G^{\mu}_{\ \nu} + \chi^{\mu}_{\ \nu}
    		= \frac{\kappa^2}{2}\stackrel{A}{T^\mu_{\ \nu}}  \ , 
    		\label{1:JC-linear-A} \\
   	\left[ \stackrel{(1)}{\delta\Sigma^{\mu}_{\ \nu}} 
   		- \frac{3}{4} \delta^\mu_\nu \stackrel{(1)}{\delta Q} \right] 
   		\Bigg|_{y=d}
    		&=& \frac{l}{2\Omega^2}G^{\mu}_{\ \nu} + \frac{l}{\Omega^2}
    		\left( \delta\phi^{|\mu}_{\ |\nu}-\delta^{\mu}_{\nu}
    		\Box\delta\phi \right)\
    		+\frac{\chi^{\mu}_{\ \nu}}{\Omega^4}
    		= -\frac{\kappa^2}{2\Omega^2}\stackrel{B}{T^\mu_{\ \nu}} \ .
    		\label{1:JC-linear-B}
\end{eqnarray}
Here, we used the relation~(\ref{rel:EM}) between 
$\stackrel{B}{\tilde{T}^\mu_{\ \nu}}$ and 
$\stackrel{B}{T^\mu_{\ \nu}}$. 
The homogeneous solution $ \chi^{\mu}_{\ \nu} $ can be eliminated from
Eqs.~(\ref{1:JC-linear-A}) and (\ref{1:JC-linear-B}) to yield
\begin{equation}
	G^{\mu}_{\ \nu} = \frac{\kappa^2}{l}
		\frac{1}{1-\Omega^2}
		\left( \stackrel{A}{T^\mu_{\ \nu}} + \Omega^2
		\stackrel{B}{T^\mu_{\ \nu}} \right) + 
		\frac{2\Omega^2}{1-\Omega^2}
		\left( \delta\phi^{|\mu}_{\ |\nu} - 
		\delta^{\mu}_{\nu}\Box\delta\phi \right) \ .
		\label{1:linear-eq} 
\end{equation}

We now introduce a linearized version of the field $\Psi$
introduced in Eq.~(\ref{1:STG-1}) by
 $\Psi = (1-\Omega^2) + 2\Omega^2 \delta\phi \equiv 
\Psi_0 + \delta\Psi$. The linearized effective equations
 can then be written as
\begin{equation}
	G^{\mu}_{\ \nu} = \frac{\kappa^2}{l\Psi_0}
		\left( \stackrel{A}{T^\mu_{\ \nu}} + (1-\Psi_0)
		\stackrel{B}{T^\mu_{\ \nu}} \right) + 
		\frac{1}{\Psi_0}
		\left( \delta\Psi^{|\mu}_{\ |\nu} - 
		\delta^{\mu}_{\nu}\Box\delta\Psi \right) \ . 
		\label{1:linear-STG-1}
\end{equation}
Eqs.~(\ref{JC:linear-A}) and (\ref{JC:linear-B}) determine
the homogeneous solution, $ \chi^{\mu}_{\ \nu} $, as
\begin{equation}
	\chi^{\mu}_{\nu} = - \frac{\kappa^2}{2}
		\frac{1-\Psi_0}{\Psi_0}
		\left( \stackrel{A}{T^\mu_{\ \nu}} + 
		\stackrel{B}{T^\mu_{\ \nu}} \right)
		-\frac{l}{2\Psi_0}
		\left( \delta\Psi^{|\mu}_{\ |\nu} - 
		\delta^{\mu}_{\nu}\Box\delta\Psi \right) \ .
		\label{linear:chi} 
\end{equation}
The traceless condition of $\chi^{\mu}_{\ \nu}$ leads to
\begin{equation}
	\Box\delta\Psi = \frac{\kappa^2}{l} 
		\frac{\stackrel{A}{T} + \stackrel{B}{T}}{2\omega +3} \ .
		\label{1:linear-STG-2}
\end{equation}
Thus we found the fact that linearizing our method leads 
to the ``linearized quasi-Brans-Dicke gravity" with the Brans-Dicke parameter,
\begin{equation}
	\omega = \frac{3}{2} \left( \frac{1}{\Omega^2}-1 \right)
  		= \frac{3}{2} \left( e^{2d/l}-1 \right) \ .
  		\label{linear:coupling}
\end{equation}

By linearizing $G^{\mu}_{\ \nu}$ in Eq.~(\ref{1:linear-eq}) and defining
\begin{equation}
	\bar{h}_{\mu\nu} = h_{\mu\nu} - \frac{1}{2}\eta_{\mu\nu}h \ ,
	\label{trace-reversed}
\end{equation}
one gets
\begin{equation}
	\frac{1}{2}\left( \bar{h}^{\alpha\mu}_{\quad|\nu\alpha}
		+ \bar{h}^{\alpha\ |\mu}_{\ \nu\ |\alpha}
		- \Box\bar{h}^{\mu}_{\ \nu} 
		- \delta^{\mu}_{\nu}\bar{h}_{\alpha\beta}
		^{\quad|\alpha\beta} \right)
		= \frac{2\Omega^2}{1-\Omega^2}
		\left( \delta\phi^{|\mu}_{\ |\nu} - 
		\delta^{\mu}_{\nu}\Box\delta\phi \right) + \frac{\kappa^2}{l}
		\frac{1}{1-\Omega^2}
		\left( \stackrel{A}{T^\mu_{\ \nu}} + \Omega^2
		\stackrel{B}{T^\mu_{\ \nu}} \right) \ .
		\label{1:pre-linear-equation}
\end{equation}
Note that $h^{\mu}_{\ \nu}(x)$ is the fluctuation of the
induced metric on the $A$-brane located at $y=0$. 
The gauge freedom can be used to set
\begin{equation}
	\bar{h}^{\alpha\mu}_{\quad|\alpha} = \frac{2\Omega^2}{1-\Omega^2}
		\delta\phi^{|\mu} \ ,
		\label{gauge1}
\end{equation}	
and then Eq.~(\ref{1:pre-linear-equation}) becomes
\begin{equation}
	\Box\bar{h}^{\mu}_{\ \nu}
		= -\frac{2\kappa^2}{l}\frac{1}{1-\Omega^2}
		\left( \stackrel{A}{T^\mu_{\ \nu}} + \Omega^2
		\stackrel{B}{T^\mu_{\ \nu}} \right) +
		\frac{\kappa^2}{3l}\frac{\Omega^2}{1-\Omega^2}
		\delta^{\mu}_{\nu}
		\left( \stackrel{A}{T} + \stackrel{B}{T} \right) \ .
		\label{1:wave}
\end{equation}
This is in agreement with the leading order term in Eq.~(\ref{eq:g-wave}) and
of course is the same as the one derived by Garriga and Tanaka~\cite{GT}.

\subsection{2-nd order}

Next we compute the 2-nd order solution. The basic equations become  
\begin{eqnarray}
 	& & \stackrel{(2)}{\delta\Sigma^\mu_{\ \nu , y}} - 
 		\frac{4}{l} \stackrel{(2)}{\delta\Sigma^\mu_{\ \nu}} 
    		= -\left[ R^{\mu}_{\ \nu} - {1\over 4} \delta^\mu_\nu R 
      		\right]^{(2)}      \ , 
      		\label{2:linear-munu-traceless}    \\
 	& &  \stackrel{(2)}{\delta Q} 
     		= \frac{l}{6}\left[ R \right]^{(2)}  \ ,  
     		\label{2:linear-munu-trace} \\
 	& &  \stackrel{(2)}{\delta Q_{, y}} -{2\over l}
 		\stackrel{(2)}{\delta Q} = 0\ ,  
 		\label{2:linear-yy} \\
 	& &  \nabla_\lambda \stackrel{(2)}{\delta\Sigma_{\mu }^{\ \lambda}}  
        	-  {3\over 4} \nabla_\mu \stackrel{(2)}{\delta Q} = 0   \ .
        	\label{2:linear-ymu}
\end{eqnarray}
The junction conditions are
\begin{eqnarray}
   	\left[ \stackrel{(2)}{\delta\Sigma^{\mu}_{\ \nu}} 
   		- \frac{3}{4} \delta^\mu_\nu \stackrel{(2)}{\delta Q} \right] 
   		\Bigg|_{y=0}
    		&=& \frac{\kappa^2}{2}\stackrel{A(2)}{T^\mu_{\ \nu}}  \ , 
    		\label{2:JC-linear-A} \\
   	\left[ \stackrel{(2)}{\delta\Sigma^{\mu}_{\ \nu}} 
   		- \frac{3}{4} \delta^\mu_\nu \stackrel{(2)}{\delta Q} \right] 
   		\Bigg|_{y=d}
    		&=& -\frac{\kappa^2}{2\Omega^2}\stackrel{B(2)}{T^\mu_{\ \nu}}
    		\ . \label{2:JC-linear-B}
\end{eqnarray}
 From Eqs.~(\ref{2:linear-munu-traceless}) and (\ref{2:linear-munu-trace}), 
the solution is
\begin{equation}
	\stackrel{(2)}{\delta Q} = 0 \ ,
	\label{2:linear-Q}
\end{equation}
and
\begin{equation}	
	\stackrel{(2)}{\delta\Sigma^{\mu}_{\ \nu}}
		= \frac{l^2}{4}\left( \frac{y}{a^4} + \frac{l}{2a^2} \right) 
		{\cal S}^{\mu}_{\ \nu}
		- \frac{l^2}{8}\left( \frac{1}{a^6} + \frac{1}{a^2} \right) 
		\Box\chi^{\mu}_{\ \nu} + \frac{C^{\mu}_{\ \nu}}{a^4}, \quad 
		C^{\mu}_{\ \mu} = 0 \ .
		\label{2:linear-sigma}
\end{equation}	
Here we have introduced the tensor ${\cal S}^\mu_\nu$,
\begin{equation}
	{\cal S}^{\mu}_{\ \nu} = \frac{1}{3}R^{|\mu}_{\ |\nu} - 
		\Box R^{\mu}_{\nu}
		+ \frac{1}{6}\delta^{\mu}_{\ \nu}\Box R, 
		\label{eq:linear-s}
\end{equation}
with the properties 
${\cal S}^{\mu}_{\ \nu|\mu} = {\cal S}^{\mu}_{\ \mu}=0$.

Eq.~(\ref{2:linear-yy}) is trivially satisfied by $\stackrel{(2)}{\delta Q}$
in Eq.~(\ref{2:linear-Q}).
 To satisfy Eq.~(\ref{2:linear-ymu}), 
the homogeneous solution in $\stackrel{(2)}{\delta\Sigma^{\mu}_{\ \nu}}$
 is constrained as $C^{\mu}_{\ \nu|\mu} = 0$. 
The junction conditions (\ref{2:JC-linear-A}) and (\ref{2:JC-linear-B})
then give
\begin{eqnarray}
   	& & \frac{l^3}{8}{\cal S}^{\mu}_{\ \nu} - \frac{l^2}{4}
    		\Box\chi^{\mu}_{\ \nu} 
    		+ C^{\mu}_{\ \nu}
    		= \frac{\kappa^2}{2}\stackrel{A(2)}{T^\mu_{\ \nu}}  \ , 
    		\label{2:linear-A} \\
   	& & \frac{l^2}{4}\left( \frac{d}{\Omega^4}+\frac{l}{2\Omega^2}
    		 \right) {\cal S}^{\mu}_{\ \nu} - 
    		\frac{l^2}{8}\left( \frac{1}{\Omega^6}+\frac{1}{\Omega^2} 
    		\right) \Box\chi^\mu_{\ \nu} + \frac{C^{\mu}_{\ \nu}}{\Omega^4}
    		= -\frac{\kappa^2}{2\Omega^2}\stackrel{B(2)}{T^\mu_{\ \nu}}
    		\ . \label{2:linear-B}
\end{eqnarray}
By combining Eqs.~(\ref{2:linear-A}) and (\ref{2:linear-B}) with the 
junction conditions at the 
first order, we obtain the following equations:
\begin{equation}
   	\frac{l}{2}G^{\mu}_{\ \nu} + \chi^{\mu}_{\ \nu} + 
   		\frac{l^3}{8}{\cal S}^{\mu}_{\ \nu} 
   		- \frac{l^2}{4}\Box\chi^{\mu}_{\ \nu} 
    		+ C^{\mu}_{\ \nu} 
    		= \frac{\kappa^2}{2}\stackrel{A}{T^\mu_{\ \nu}}  \ ,
\end{equation}    
\begin{equation}    
   	\frac{l}{2}G^{\mu}_{\ \nu} + \frac{\chi^{\mu}_{\ \nu}}{\Omega^2} +
   	\frac{l^2}{4}\left( \frac{d}{\Omega^2}+\frac{l}{2} \right) 
    	{\cal S}^{\mu}_{\ \nu} - 
    	\frac{l^2}{8}\left( \frac{1}{\Omega^4}+ 1  \right) 
    	\Box\chi^\mu_{\ \nu} + l\left( \delta\phi^{|\mu}_{\ |\nu} - 
    	\delta^{\mu}_{\nu}\Box\delta\phi \right) + 
    	\frac{C^{\mu}_{\ \nu}}{\Omega^2}
    	= -\frac{\kappa^2}{2}\stackrel{B}{T^\mu_{\ \nu}} \ .
\end{equation}

Eliminating $C^{\mu}_{\ \nu}$ from the above
two equations, we obtain
the effective 4-dimensional theory of gravity with correction terms;
\begin{equation}
	G^{\mu}_{\ \nu} = \frac{\kappa^2}{l}\frac{1}{1-\Omega^2}
	\left( \stackrel{A}{T^\mu_{\ \nu}} + \Omega^2
	\stackrel{B}{T^\mu_{\ \nu}} \right) + 
	\frac{2\Omega^2}{1-\Omega^2}
	\left( \delta\phi^{|\mu}_{\ |\nu} - 
	\delta^{\mu}_{\nu}\Box\delta\phi \right) 
	+ \frac{l^2}{2(1-\Omega^2)}
	\left( \frac{d}{l}-\frac{1-\Omega^2}{2} \right) {\cal S}^{\mu}_{\ \nu}
	- \frac{l}{4}\frac{1-\Omega^2}{\Omega^2}\Box\chi^{\mu}_{\ \nu} \ .
	\label{2:QBD}
\end{equation}
If we rewrite this equation using $\Psi$,  we get
\begin{equation}
	G^{\mu}_{\ \nu} = \frac{\kappa^2}{l\Psi_0}
	\left( \stackrel{A}{T^\mu_{\ \nu}} + (1-\Psi_0)
	\stackrel{B}{T^\mu_{\ \nu}} \right) + 
	\frac{1}{\Psi_0}
	\left( \delta\Psi^{|\mu}_{\ |\nu} - 
	\delta^{\mu}_{\nu}\Box\delta\Psi \right) 
	+ \frac{l^2}{2\Psi_0}
	\left( \frac{d}{l}-\frac{\Psi_0}{2} \right) {\cal S}^{\mu}_{\ \nu}
	- \frac{l}{4}\frac{\Psi_0}{1-\Psi_0}\Box\chi^{\mu}_{\ \nu} \ .
\end{equation}
Note that $\chi^{\mu}_{\ \nu}$, given by the
traceless part of Eq.~(\ref{linear:chi}), satisfies the 
transverse-traceless condition. Thus we have obtained the lenearized 
quasi-Brans-Dicke theory with Kaluza-Klein corrections.
Using $\bar{h}_{\mu\nu}$ defined by Eq.~(\ref{trace-reversed}), 
Eq.~(\ref{2:QBD}) leads to
\begin{eqnarray}
	& & \frac{1}{2} \left( \bar{h}^{\alpha\mu}_{\quad|\nu\alpha}
		+ \bar{h}^{\alpha\ |\mu}_{\ \nu\ |\alpha}
		- \Box\bar{h}^{\mu}_{\ \nu} 
		- \delta^{\mu}_{\nu}\bar{h}_{\alpha\beta}
		^{\quad|\alpha\beta} \right)  \nonumber \\
		& & \quad = \frac{\kappa^2}{l}\frac{1}{1-\Omega^2}
		\left( \stackrel{A}{T^\mu_{\ \nu}} + \Omega^2
		\stackrel{B}{T^\mu_{\ \nu}} \right) 
		+ \frac{2\Omega^2}{1-\Omega^2}
		\left( \delta\phi^{|\mu}_{\ |\nu} - 
		\delta^{\mu}_{\nu}\Box\delta\phi \right) \nonumber \\
		& & \quad + \frac{l^2}{2(1-\Omega^2)}\left( \frac{d}{l}-
		\frac{1-\Omega^2}{2} \right)
		\left(
		\frac{1}{3}\bar{h}^{\alpha\beta\quad|\mu}
		_{\quad|\alpha\beta\ |\nu}
		+\frac{1}{6}\Box\bar{h}^{|\mu}_{\ |\nu} - 
		\frac{1}{2}\Box\bar{h}^{\mu\alpha}_{\quad|\nu\alpha} -
		\frac{1}{2}\Box\bar{h}^{\alpha\ |\mu}_{\ |\nu\ |\alpha} 
		+\frac{1}{2}\Box^2\bar{h}^{\mu}_{\ \nu} +
		\frac{1}{6}\delta^{\mu}_{\nu}\Box\bar{h}_{\alpha\beta}^
		{\quad|\alpha\beta} -
		\frac{1}{6}\delta^{\mu}_{\nu}\Box^2 \bar{h}	 
		\right) \nonumber \\
		& & \quad + \frac{l}{4}\frac{1-\Omega^2}{\Omega^2}
		\left[ \frac{l\Omega^2}{1-\Omega^2} 
		\left( \Box\delta\phi^{|\mu}_{\ |\nu} - 
		\delta^{\mu}_{\nu}\Box^2\delta\phi \right) -
		\frac{\kappa^2}{2}\frac{\Omega^2}{1-\Omega^2}\Box
		\left(\stackrel{A}{T^\mu_{\ \nu}} + 
		\stackrel{B}{T^\mu_{\ \nu}} \right) \right] \ .
\end{eqnarray}
Imposing the gauge condition,
\begin{equation}
	\bar{h}^{\alpha\mu}_{\quad|\alpha} = \frac{2\Omega^2}{1-\Omega^2}
		\delta\phi^{|\mu} - \left[
		\frac{4l^2}{3}\frac{\Omega^2}{1-\Omega^2}
		\left( \frac{d/l}{2(1-\Omega^2)}-\frac{1}{4} \right) 
		- \frac{l^2}{4}
		\right] \Box\delta\phi^{|\mu} + 
		\frac{l^2}{6}\left[ \frac{d/l}{2(1-\Omega^2)}-
		\frac{1}{4} \right]\Box\bar{h}^{|\mu} \ ,
\end{equation}	
we get the following equation,
\begin{eqnarray}
	\Box\bar{h}^{\mu}_{\ \nu}
		&=& -\frac{2\kappa^2}{l}\frac{1}{1-\Omega^2}
		\left( \stackrel{A}{T^\mu_{\ \nu}} + \Omega^2
		\stackrel{B}{T^\mu_{\ \nu}} \right) +
		\frac{\kappa^2}{3l}\frac{\Omega^2}{1-\Omega^2}
		\delta^{\mu}_{\nu}
		\left( \stackrel{A}{T} + \stackrel{B}{T} \right) \nonumber \\
		& & -\frac{2l\kappa^2}{1-\Omega^2}
		\left[ \frac{3}{8}-\frac{1}{8}\Omega^2 
		- \frac{d/l}{2(1-\Omega^2)}
		\right] \Box \left( \stackrel{A}{T^\mu_{\ \nu}} - \frac{1}{6}
		\delta^{\mu}_{\nu}\stackrel{A}{T} \right) \nonumber \\
		& & - \frac{2l\kappa^2\Omega^2}{1-\Omega^2}
		\left[ \frac{1}{8}+\frac{1}{8}\frac{1}{\Omega^2} - 
		\frac{d/l}{2(1-\Omega^2)}\right] 
		\Box \left( \stackrel{B}{T^\mu_{\ \nu}} - \frac{1}{6}
		\delta^{\mu}_{\nu}\stackrel{B}{T} \right) \ .
\end{eqnarray}
This result coincides with the result of the standard
linear theory given in Eq.~(\ref{eq:g-wave}).


\end{document}